\documentclass[]{aa}

\usepackage[T1]{fontenc}
\usepackage[normalem]{ulem}
\usepackage{graphicx}
\usepackage{natbib}
\bibpunct{(}{)}{;}{a}{}{,}
\usepackage{scrextend}		
\usepackage{subcaption}
\usepackage{multirow}
\usepackage{amssymb}	
\usepackage{xfrac} 
\usepackage{dsfont}
\usepackage{mwe,tikz}
\usepackage[percent]{overpic}
\usepackage{lipsum}
\usepackage{siunitx}
\usepackage{rotating}
\usepackage{xspace}
\usepackage{savesym}
\usepackage{amsmath}
\savesymbol{iint}
\usepackage{txfonts}
\restoresymbol{TXF}{iint}
\usepackage{hyperref}

\usepackage{longtable}
\usepackage{supertabular,booktabs}

\newcommand{\mdotin}{$\dot{m}_{in}$\xspace}

\newcommand{\rj}{$r_J$\xspace}
\newcommand{\mdotint}{$\dot{m}_{in} (t)$\xspace}

\newcommand{\rjt}{$r_J (t)$\xspace}

\newcommand{\pair}{$(r_J, \dot{m}_{in})$\xspace}

\newcommand{\ie}{i.e.\xspace}
\newcommand{\eg}{e.g.\xspace}

\newcommand{\gx}{GX\,339-4\xspace}
\newcommand{\jed}{jet-emitting disk\xspace}

\newcommand{\sad}{standard accretion disk\xspace}

\newcommand{\HT}{hard tail\xspace}

\newcommand{\one}{$\# 1$\xspace}
\newcommand{\two}{$\# 2$\xspace}
\newcommand{\three}{$\# 3$\xspace}
\newcommand{\four}{$\# 4$\xspace}

\newcommand{\tA}{Type A\xspace}
\newcommand{\tB}{Type B\xspace}
\newcommand{\tC}{Type C\xspace}

\newcommand{\newA}{0\xspace}
\newcommand{\newB}{3\xspace}
\newcommand{\newC}{4\xspace}

\begin{document} 

 \title{A unified accretion-ejection paradigm for black hole X-ray binaries}

 \subtitle{V. Low-frequency quasi-periodic oscillations}

 \author{G. Marcel \inst{1}
  \and
  F. Cangemi\inst{2}
  \and
  J. Rodriguez\inst{2}
  \and
  J. Neilsen\inst{1}
  \and
  J. Ferreira\inst{3}
  \and
  P.-O. Petrucci\inst{3}
  \and
  J. Malzac\inst{4}
  \and
  S. Barnier\inst{3}
  \and
  M. Clavel\inst{3}
  }
  
 \institute{Villanova University, Department of Physics, Villanova, PA 19085, USA \\
   \email{gregoire.marcel@villanova.edu or gregoiremarcel26@gmail.com}
   \and
   AIM, CEA, CNRS, Université Paris-Saclay, Université Paris Diderot, Sorbonne Paris Cité, F-91191 Gif-sur-Yvette, France
   \and
   Univ. Grenoble Alpes, CNRS, IPAG, 38000 Grenoble, France
   \and 
   IRAP, Université de Toulouse, CNRS, UPS, CNES, Toulouse, France
   }

 \date{Received 21 January, 2020; accepted 22 May, 2020}

 \abstract
 {We proposed that the spectral evolution of transient X-ray binaries (XrBs) is due to an interplay between two flows: a standard accretion disk (SAD) in the outer parts and a jet-emitting disk (JED) in the inner parts. We showed that the spectral evolution in X-ray and radio during the 2010-2011 outburst of \gx can be recovered. However, the observed variability in the X-ray wavelength was never addressed in this framework.}%
 {We investigate the presence of low-frequency quasi-periodic oscillations (LFQPOs) during an X-ray outburst, and address the possible correlation between the frequencies of these LFQPOs and the transition radius between the two flows, \rj.}%
 {We selected X-ray and radio data that correspond to three outbursts of \gx. We used the method detailed in previous papers to obtain the best parameters, \rjt and \mdotint, for each outburst. We also independently searched for X-ray QPOs in each selected spectra and compared the QPO frequency to the Kepler and epicyclic frequencies of the flow in \rj.}%
 {We successfully reproduce the evolution of the X-ray spectra and the radio emission for three different activity cycles of \gx . We use a unique normalization factor for the radio emission, $\tilde{f}_R$. We also report the detection of seven new LFQPOs (three \tB and four \tC ), in addition to those previously reported in the literature. We show that the \tC QPOs frequency can be linked to the dynamical JED-SAD transition radius \rj, rather than to the optically thin-thick transition radius in the disk. The scaling factor $q$ such that $\nu_{QPO} \simeq \nu_K (r_J) / q$ is $q \simeq 70-140$; this factor is consistent over the four cycles and is similar to previous studies.}%
 {The JED-SAD hybrid disk configuration not only provides a successful paradigm that allows us to describe XrB cycles, but also matches the evolution of QPO frequencies. \tC QPOs provide an indirect way to probe the JED-SAD transition radius, where an undetermined process produces secular variability. The demonstrated relation between the transition radius links \tC QPOs to the transition between two different flows, effectively tying it to the inner magnetized structure, that is, the jets. This direct connection between the (accretion-ejection) structure of the jets and the process responsible for \tC QPOs, if confirmed, could naturally explain their puzzling multiwavelength behavior.}%


 \maketitle
%


\section{Introduction}

The generic behavior of X-ray binaries (XrBs) is now adequately captured in the literature \citep[\eg, ][]{Dunn10}. These systems spend most of their lives in a quiescent and barely detectable state, sometimes for years, before undergoing sudden X-ray outbursts lasting from weeks to months. These outbursts are accompanied by spectral changes following a similar pattern for most objects. Starting from quiescence, the total luminosity increases in both X-ray and radio bands. Radio flux is detectable and the X-ray spectrum peaks above $10\,$keV: this state is called the hard state. Systems remain in this state over several orders of magnitude in X-ray and radio luminosities. At some point, the radio flux vanishes and the X-ray spectrum then peaks around $1\,$keV: this is the soft state\footnote{We would like to point out that in some cases the system never reaches the soft state before the luminosity decreases back to quiescence: such outbursts are referred to as 'failed', or 'hard-only', outbursts \citep{Tetarenko16}.}. Once in the soft state, the flux eventually decreases until the source transitions back to the hard state, along with the reappearance of detectable radio fluxes. In the so-called hardness-intensity diagram \citep{Kording06}, this behavior produces the archetypal \textquotesingle q\textquotesingle -cycle of XrBs \citep[for a review, see for example][]{Dunn10}. To date, there is no consensus regarding an explanation of these cycles \citep{Remillard06, Yuan14}. It is however very likely that the X-ray spectral changes are due to variations in the inner accretion flow structure \citep[see for example][and references therein]{Done07}.

To explain this behavior, \citet{Esin97}\footnote{See also, for example, \citet{Thorne75}, \citet{Shapiro76}, \citet{Oda77}, \citet{Abramowicz80}, or \citet{Lasota96}} envision the interplay between an outer \sad \citep[SAD hereafter;][]{SS73} and an inner advection-dominated flow \citep{Ichimaru77, Rees82, Narayan94}. Although the presence of a SAD in the outer regions seems inevitable \citep{Done07}, the inner advection-dominated flow structure remains uncertain. The many scenarios following \citet{Esin97} notably fail to explain the radio (non-)detections. Radio detections are commonly interpreted as persistent self-collimated jets \citep{BK79, Mirabel92}, whereas non-detections in radio are the result of jet quenching (\citeauthor{Corbel04} \citeyear{Corbel04}, \citeauthor{Fender04} \citeyear{Fender04}, see however \citeauthor{Drappeau17} \citeyear{Drappeau17}, for an alternative view). Ignoring the formation and quenching of jets leaves important observational diagnostics unexplained \citep[for recent discussions, see][]{Yuan14, Marcel18a}.

A framework addressing the full accretion-ejection phenomenon was proposed and progressively elaborated on in a series of papers. \citet{Ferreira06}, hereafter paper~I, proposed that the disk be threaded by a large-scale vertical magnetic field $B_z$, which was built up mostly by accumulation from the outer disk regions. In this configuration the magnetization $\mu= B_z^2/P$, where $P$ is the total (gas plus radiation) pressure at the disk midplane, increases inwardly to reach an expected threshold value $\mu \sim 0.5$. A \jed (hereafter JED) emerges. While JED can nicely reproduce bright hard states at luminosity levels never achieved in any other accretion model \citep{Yuan14}, a sole JED configuration cannot explain the spectral cycles as those of \gx \citep[][hereafter paper~II]{Marcel18a}. When transitioning to the soft state, the system needs to not only emit a sufficiently soft spectrum, but also to fully quench its jets. Similar to \citet{Esin97}, we imagined the existence of a transition at some radius \rj, from an inner JED to an outer SAD, as already proposed in paper~I. We note that very recent numerical simulations naturally show such a magnetic field distribution: $\mu \gtrsim 0.1$ in the inner region and $\mu \ll 1$ in the outer region \citep{Scepi19, Liska19}. In \citet{Marcel18b}, hereafter paper~III, we showed that the observed domain in X-ray luminosities and hardness ratios during a standard transient X-ray binary (XrB) cycles can be covered by changing \rj and the inner accretion rate $\dot{m} (r_{isco}) = \dot{m}_{in}$. Along with these X-ray signatures, JED-SAD configurations naturally account for the radio emission whenever it is observed. As an illustration, we successfully reproduced five canonical spectral states (X-ray+radio) typically observed along a cycle. In \citet{Marcel19}, hereafter paper~IV, we independently reproduced each step of the spectral evolution of the 2010--2011 outburst from \gx, using $35$ observed radio fluxes and $297$ X-ray spectral fits from \citet{Clavel16}. We showed that a smooth evolution in disk accretion rate and transition radius can simultaneously reproduce the behavior of \gx in the X-ray and radio bands. For the first time, a time-evolution of physical parameters reproduced the behavior of a given X-ray binary in multiple spectral bands. However, there are multiple open questions remaining, such as the origin of timing properties in the JED-SAD paradigm.

In particular, quasi-periodic oscillations (QPOs) of the X-ray flux are ubiquitous features of XrBs \citep[see, e.g.,][and references therein]{1977SSRv...20..687M,Samimi79,Zhang13}. When studying the power density spectra (PDS) in Fourier space \citep{1989ARA&A..27..517V}, peaks are observed at varying frequencies and with varying widths \citep{Miyamoto91,2005ApJ...623..383H}. These peaks, called QPOs, have been detected in a very wide number of XrBs \citep{Zhang13}. QPOs are observed to evolve with the X-ray spectral shape, but remain a considerable unknown of the behavior of XrBs \citep[see][for a recent review]{Motta16}. They cover a wide range of frequencies up to kHz, but we focus in this paper on low-frequency ($0.1-10\,$Hz) QPOs. The three types of QPOs, A, B, and C, are defined using their frequencies, width, broadband noise, and amplitude \citep{2005ApJ...629..403C}. \tA QPOs are characterized by a weak and broad peak around $6-8\,$Hz, the absence of broadband noise, and are found on the soft side of state transitions (hard $\leftrightarrow$ soft). They are the rarest type of low-frequency QPOs. \tB QPOs are characterized by an intermediate and narrow peak, varying between $1-6\,$Hz, the absence of broadband noise, and are detected during the soft-intermediate state. Their detection is much more common than \tA, but most detected QPOs are of the next type. \tC are characterized by a strong and very narrow peak, varying between $0.1-10\,$Hz, a very important broadband noise, and are detected during the hard and hard-intermediate states. \tC are, by far, the most studied type of QPO in the literature.
We discerned three key features of \tC LFQPOs from observations \citep{Remillard06,Motta16}. First, the stability and persistence of the frequencies suggest that \tC LFQPOs originate in the accretion flow itself. Second, they vary significantly in frequency, especially during state transitions, that is, when the accretion flow structure is expected to change the most. Third, their root mean square (rms) amplitude is strongest in the hard X-ray band, which indicates a connection with the power-law component of the X-ray spectrum. Guided by these important properties, LFQPOs are commonly associated with the inner hot flow \citep[see section~4.4.1 in][]{McClintock06}. Indeed, a link between the LFQPOs and the outer radius of the inner flow was observed in very early works \citep[\eg, ][]{Muno99, 2000ApJ...531..537S, 1999AstL...25..718T, 2000MNRAS.312..151R, Rodriguez02, 2004ApJ...612.1018R}. 

In this paper, we intend to investigate a possible correlation between all types of QPO frequencies and the JED parameters. In section \ref{sec:2} we present the selected data sets in both X-ray and radio bands, our method of obtaining estimates of \rj and \mdotin , and the fitting results applying the method from paper~IV. Later, in the same section, we also present the methodology and results concerning the LFQPOs, and compare these to the literature. In section \ref{sec:linkQPOs}, we investigate the link between the frequency of the QPOs obtained/selected and the accretion flow structure. We finish by some discussions, conclusions, and upcoming work in section \ref{sec:Ccl}.

\begin{figure*}[t!]
 \includegraphics[width=1.0\linewidth]{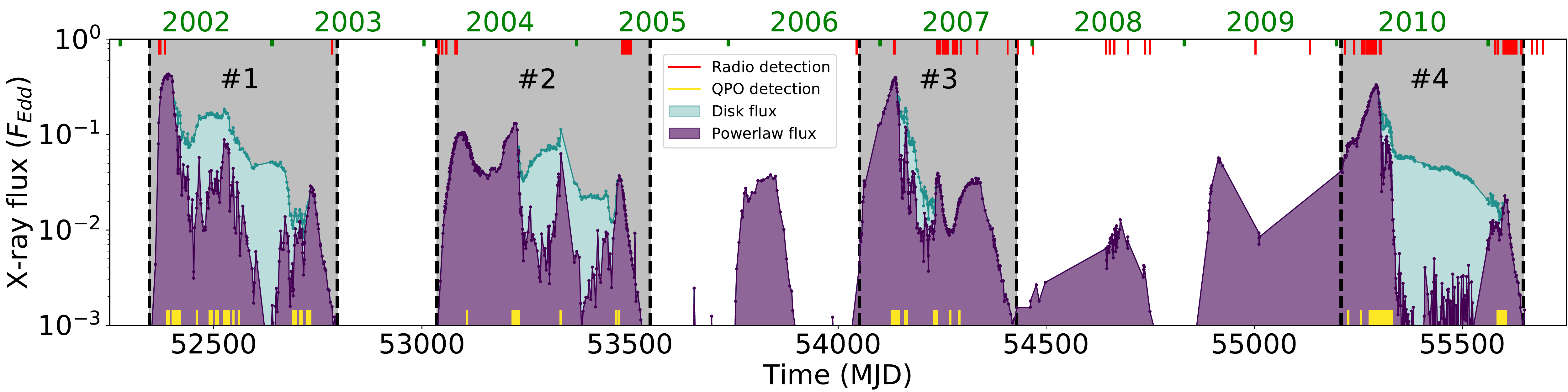}
 \caption{Light curves from \textit{RTXE}/PCA of \gx in the $3-200\,$keV energy band from 2002 to 2011 (see upper X-axis). The power law (violet) and disk (cyan) unabsorbed fluxes from the \citet{Clavel16} fits are shown. The area corresponding to the four complete outbursts (\one , \two , \three , \four ) is indicated in gray. The red lines at the top correspond to dates when steady radio fluxes were observed with the ATCA at $9\,$GHz \citep{2013MNRAS.431L.107C, Corbel13}. The yellow lines at the bottom show previous detections of QPOs \citep[][]{Motta11, Nandi12, Gao14, 2017ApJ...845..143Z}.}
 \label{fig:4outbursts}
\end{figure*}


\section{Multispectral reproduction of outbursts} \label{sec:2}

In this section, we summarize the procedure we followed to reproduce the observed spectra within our theoretical framework. The caveats, statistical and systematic errors, and limitations of the results were extensively discussed in paper~IV, and we will only develop the major outcomes here.

\subsection{Data selection} \label{sec:data}

We selected the archetypal object \gx to investigate the capability of our theoretical model to reproduce XrBs in outbursts. We chose this object because of its historic place and its short recurrence time of outbursts, once every two years on average \citep{Tetarenko16}. We used two different data sets: X-ray spectra and radio fluxes. In X-ray, we selected the proportional counter array (PCA) $3-40\,$keV data from the Rossi X-ray timing explorer (\textit{RXTE}), already reduced and fitted by \citet{Clavel16}. In radio, we used results obtained with the Australia telescope compact array (ATCA) \citep{2013MNRAS.431L.107C, Corbel13}.

We show in Fig.~\ref{fig:4outbursts} the unabsorbed light curves\footnote{The fluxes were extrapolated over the $3-200\,$keV range, see section~3.1 in paper~IV.} of the two additive models from \citet{Clavel16} fits: the power law and disk. We show the dates when LFQPOs were detected by previous studies in yellow markers on the lower X-axis \citep{Motta11, Nandi12, Gao14, 2017ApJ...845..143Z}, and when the source was observed and detected in radio in red markers on the upper X-axis \citep{Corbel13}. We can already note that the coverage of radio observations is extremely diverse: while the 2010--2011 outburst is widely covered, the 2002--2003 region was only observed at two very different stages. Four zones are highlighted in gray in this figure, defining major (and full) outbursts. The exact dates of these events are reported in Table~\ref{table:GX}.

\begin{table}[h!]
\centering
\caption{Four full cycles from \gx during the 2001--2011 decade and their number of observations in X-rays and radio.}
\begin{tabular}{c | c c c c c c}
\hline outburst & start & end & $\#$ of X-ray & $\#$ of radio \\ 
$\#$ & $(MJD)$ & $(MJD)$ & observations & detections \\ \hline
\textbf{1} & 52345 & 52796 & 212 & 5 \\
\textbf{2} & 53036 & 53548 & 277 & 16 \\
\textbf{3} & 54052 & 54429 & 250 & 15 \\ 
\textbf{4} & 55208 & 55656 & 297 & 35 \\ \hline
\end{tabular}
\label{table:GX} 
\end{table}

Outburst \four is the 2010--2011 outburst previously reproduced in paper~IV, whereas outbursts \one, \two, and \three correspond to 2002--2003, 2004--2005, and 2006--2007 respectively. We note that all these outbursts follow the standard trend described in the introduction: an increase in luminosity, the appearance of a disk component, a decrease in luminosity. There are also hard-only (failed) outbursts in 2006 and 2008--2009, which consist of luminosity increases in which the disk component is never detected. We define spectral states solely based on the shape of the continuum, \ie no timing or radio properties.

\subsection{Methodology} \label{sec:proc}

We chose the following global parameters. (i) Source distance $d \simeq 8 \pm 1\,$kpc \citep{Hynes04, Zdz04, Parker16}. (ii) Black hole mass $m = M/M_{\odot} = 5.8$, where $M_{\odot}$ is the mass of the Sun \citep[][]{Hynes03,Munoz08,Parker16,Heida17}. (iii) Disk innermost stable circular orbit\footnote{In the previous papers of this series, we used a notation $r_{in}$ for $r_{isco}$. We however decided to use only $r_{isco}$ now to avoid any ambiguity with the inner radius of other models, often labeled $r_{int}$.} $r_{isco} = r_{in} = R_{in} / R_g = 2$ \citep[\ie, spin 0.94;][]{Reis08, Miller08, Garcia15}, where $R_g = GM/c^2$ is the gravitational radius, $G$ the gravitational constant, $c$ the speed of light, and $M$ the black hole mass.
In this work, the disk accretion rate is normalized with respect to the Eddington rate $\dot{m} = \dot{M} / \dot{M}_{Edd} = \dot{M} c^2 / L_{Edd}$ \citep{Eddington}. We note that this definition of $\dot{m}$ does not include any accretion efficiency. In practice, we mostly use the accretion rate at the innermost disk radius $\dot{m}_{in}=\dot{m}(r_{isco})$, and we recall that we assume $\dot{m} (r) = \dot{m}_{in} (r/r_{isco})^{0.01}$ in the JED region (paper~II). 

We simulated a large set of parameters $r_J \in [r_{isco}=2,\,10^3]$ and $\dot{m}_{in} \in [10^{-3},\,10]$. For each pair \pair, we 
self-consistently computed the thermal balance of the hybrid disk configuration and its associated global spectrum. We then fit each simulated spectrum with the same spectral model components as those used for the spectral analysis of the observations. As a consequence, the simulated parameters of the fits can be directly compared to the observational parameters, that is, those from observational fits \citep{Clavel16}. Among these parameters, we selected the $3-200\,$keV luminosity $L_{3-200}$; the power-law fraction $PLf = L_{pl} / L_{3-200}$, where $L_{pl}$ is the power-law flux in the $3-200\,$keV energy band, and the power-law photon index $\Gamma$. This selection provides us with three constraints for any given simulated X-ray spectrum to reproduce. Additionally, we estimated for any couple \pair the synchrotron emission radiated by the jet using a self-similar approach \citep[][Appendix A in paper III, section 3.2.2 in paper IV]{BK79, Heinz03}. We can thus uniquely link the steady radio flux observed to the accretion flow structure,
\begin{eqnarray}
F_R = \tilde{f}_R \dot{m}_{in}^{17/12} r_{isco} (r_J - r_{isco})^{5/6} \, F_{Edd} \label{eq:Fr}
,\end{eqnarray}
\noindent where $\tilde{f}_R$ is a scaling factor that is found to best fit observations at $\tilde{f}_R = \num{5e-10}$ for outburst \four, and $F_{Edd} = L_{Edd}/(\nu_R 4 \pi d^2)$ the Eddington flux at $\nu_R = 9\,$Hz received at a distance $d$.

\begin{figure*}[t!]
\begin{center}
 \includegraphics[width=.99\linewidth]{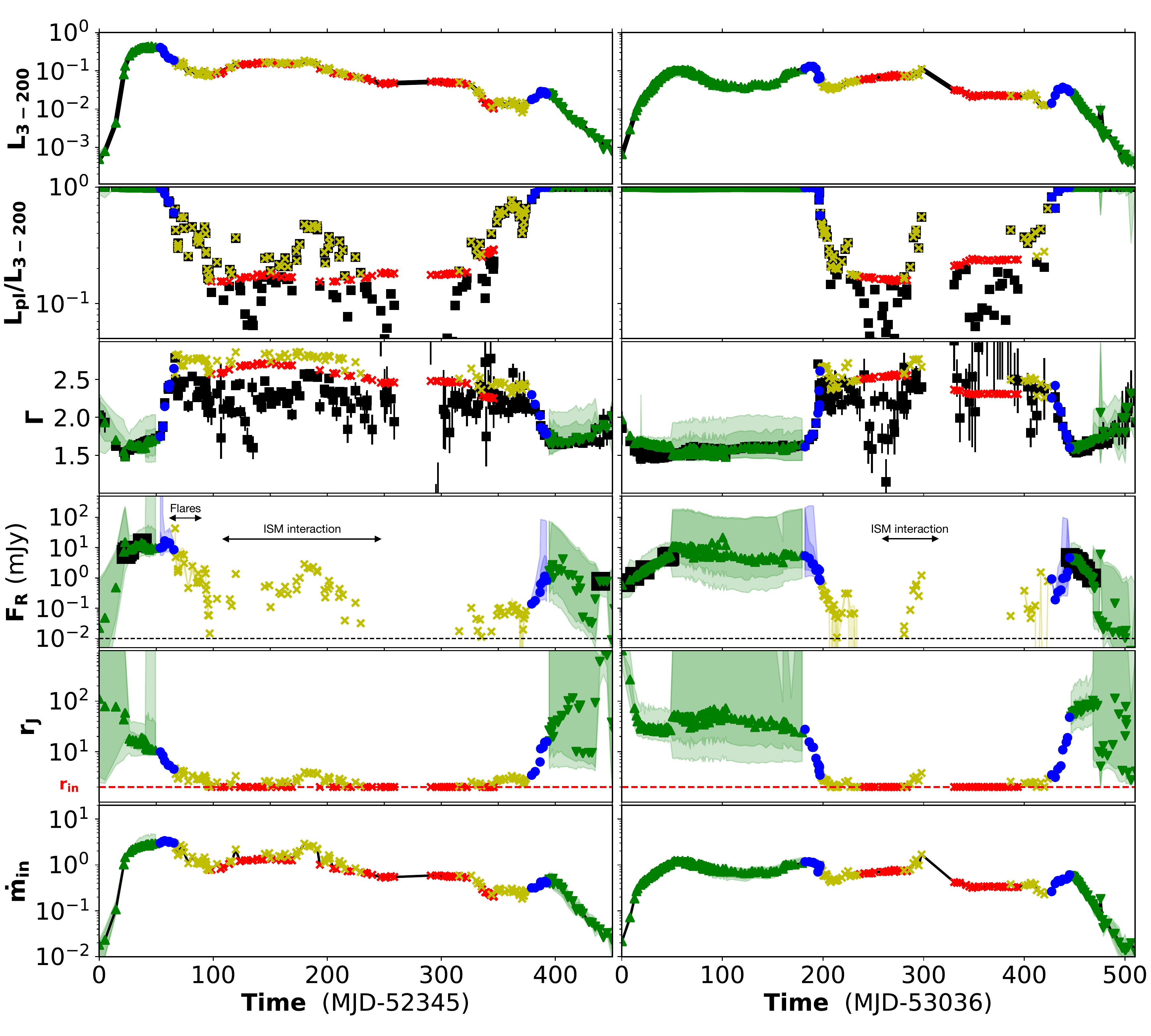}
 \caption{ Observation and model parameters for cycles \one (2002--2003, left) and \two (2004--2005, right). From top to bottom: X-ray flux and power-law fraction in the $3-200\,$keV range, power-law index, $9\,$GHz radio flux, transition radius \rj, and inner accretion rate \mdotin (at ISCO). For the first 4 panels, \ie , the constraints, the black squares represent observations. Each figure uses the same color-code: Green upward (downward) triangles indicate the rising (decaying) hard, blue circles indicate hard-intermediate, yellow crosses show soft-intermediate, and red crosses show soft states. Additionally, in the state-associated colors we draw the $5\%$ and $10\%$ confidence intervals (\ie , $5\%$ and $10\%$ bigger $\zeta_{X+R}$, see paper~IV). Double arrows are drawn when radio emission was also observed but is interpreted as radio flares or interactions with the interstellar medium (Corbel et al., in prep.)}
 \label{fig:OB12}
 \end{center}
\end{figure*}

\begin{figure*}[t!]
\begin{center}
 \includegraphics[width=.99\linewidth]{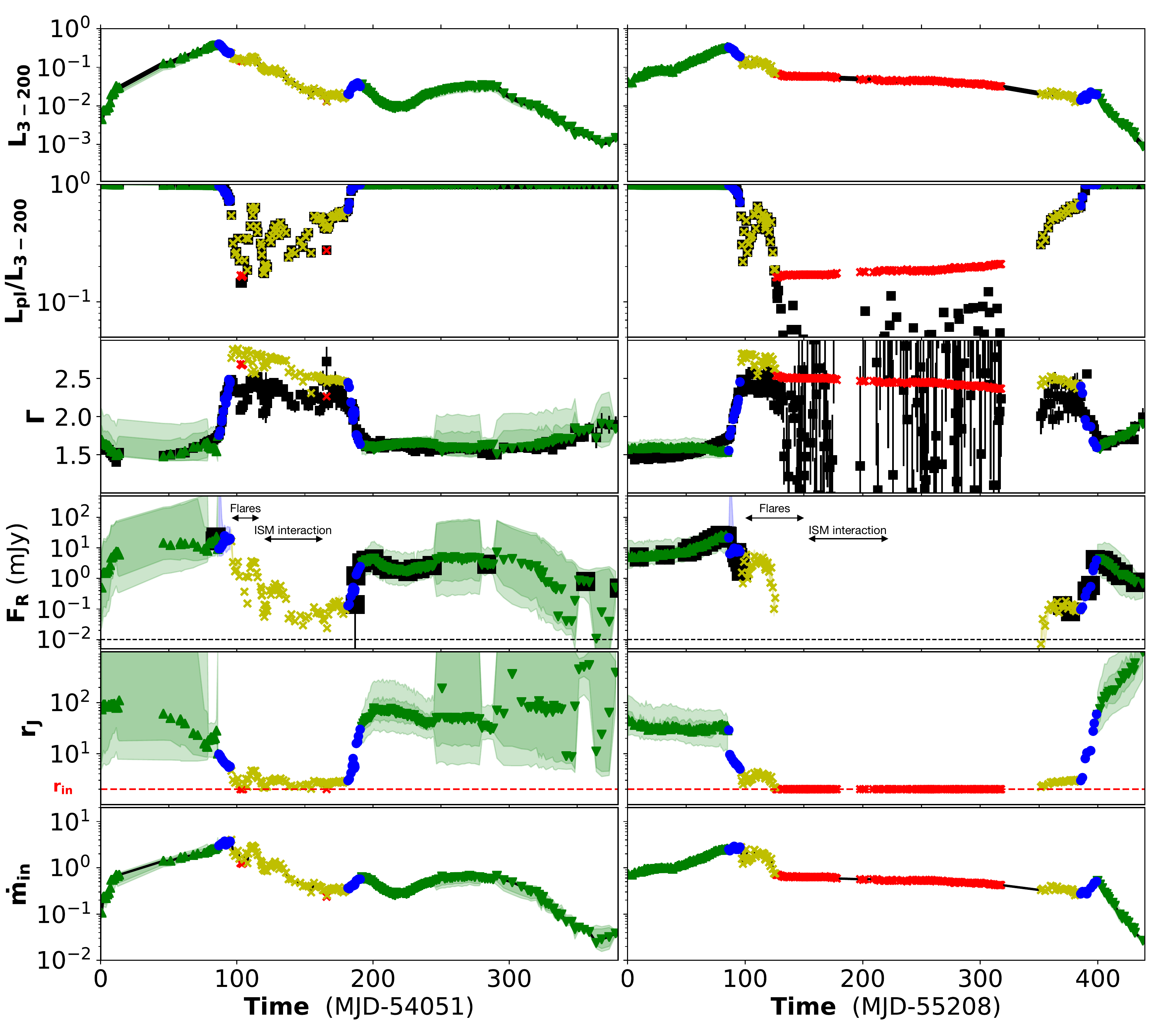}
 \caption{ Observation and model parameters for cycles \three (2006--2007, left) and \four (2010--2011, right). From top to bottom: X-ray flux and power-law fraction in the $3-200\,$keV range, power-law index, $9\,$GHz radio flux, transition radius \rj, and inner accretion rate \mdotin (at ISCO). For the first 4 panels, \ie , the constraints, the black squares represent observations. Each figure uses the same color-code: Green upward (downward) triangles indicate rising (decaying) hard, blue circles indicate hard-intermediate, yellow crosses show soft-intermediate, and red crosses show soft states. Additionally, in the state-associated colors we draw the $5\%$ and $10\%$ confidence intervals (\ie , $5\%$ and $10\%$ bigger $\zeta_{X+R}$, see paper~IV). Double arrows are drawn when radio emission was also observed but is interpreted as radio flares or interactions with the interstellar medium (Corbel et al., in prep.)}
 \label{fig:OB34}
 \end{center}
\end{figure*}

We tested different minimization procedures in paper~IV and selected the most promising for the 2010--2011 outburst. For any given observation, our procedure uses variable weights to minimize the differences with the four constraints (or parameters): the first three in X-rays ($L_{3-200}$, $PLf$, $\Gamma$), and the last one in radio ($F_{R}$). In this paper, we thus use the exact same procedure on the other three selected outbursts. The procedure searches for the pair of parameters \pair that minimizes, for each individual observation, the following function:
\begin{align}
\zeta_{X+R} =& \frac{ | \text{log} \left[ L_{3-200} / L_{3-200}^{obs} \right] | }{\alpha_{flux}} + \frac{ | \text{log} \left[ PLf / PLf_{obs} \right] | }{\alpha_{PLf}} \nonumber \\
&\hspace{1.5cm} + \frac{ | \Gamma - \Gamma^{obs} | }{\alpha_{\Gamma}} + \frac{ | \text{log} \left[ F_R / F_R^{obs} \right] | }{\alpha_{R}} \label{eq:fitproc},
\end{align}
\noindent where $L_{3-200}^{obs}$, $PLf^{obs}$, $\Gamma^{obs}$, and $F_R^{obs}$ are the observational constraints \citep{2013MNRAS.431L.107C, Corbel13, Clavel16}. The selected weights are $\alpha_{\Gamma} = 2-6 \, \text{log}_{10} (PLf)$, $\alpha_{flux} = \alpha_{PLf} = 1$, and $\alpha_{R} = 5$ (see paper~IV). When no radio flux can be estimated (\ie , during the entire soft and soft-intermediate states), we use $1/\alpha_R= 0$. 

\subsection{Results} \label{sec:res}

We use this procedure to derive the best \pair for all observations in the four outbursts. We show these results in Fig.~\ref{fig:OB12} for outbursts \one and \two and Fig.~\ref{fig:OB34} for outbursts \three and \four \footnote{The figure for outburst \four is reported for comparison, but is the same as Fig.~7 in paper~IV.}. Remarkably, the same factor $\tilde{f}_R = \num{5e-10}$ can effectively be used to reproduce all four outbursts. This parameter is a combination of many different physical properties (\eg, $B_z$ or $r_{isco}$, see papers~III, IV), but it is a simple proxy for the jet radiative efficiency. It is thus noteworthy that $\tilde{f}_R$ could remain constant all along the evolution of the four outbursts, \ie , after the quenching and building of three jets.


In the hard and hard-intermediate cases, both the X-ray flux and power-law fractions (top two panels) are very well reproduced in all four outbursts. 
The theoretical power-law index during these two states, that is, when the power law is dominant, follows observations accordingly. The model also reproduces the radio flux excellently when present, but occasionally at the expense of matching $\Gamma$. This is exemplified in the first 50 days of outburst \two, when the radio constraints are associated with small inaccuracies in $\Gamma$. However, the maximum difference is $|\Gamma_{\mathrm{th}} - \Gamma_{\mathrm{obs}}| \simeq 0.1$, that is, only twice the average error $\Delta \Gamma_{\mathrm{obs}} = 0.05$ in hard and hard-intermediate states in observational fits from \citet{Clavel16}. These differences can be corrected by local and physical parameters of our model such as the illumination fraction (paper~III) or by including reflection in the theoretical model. We would like to point out that reflection is not expected to have a huge impact, as we compare the theoretical continuum to the observed continuum extracted from observational fits that were performed using multiple reflection models \citep{Clavel16}.
We also recall that the predicted radio flux from Eq.~(\ref{eq:Fr}) is a very simple and first order approximation.

Modeling is more complex in the case of soft-intermediate and soft states. In disk-dominated states, the hard part of the X-ray spectrum is dominated by the so-called \HT, a steep power law ($\Gamma \sim 2.5$) with no clear high-energy cutoff \citep[see][]{Remillard06}. There is no consensus about the physical origin of this component, and we thus decide to use a proxy: we add a steep power law with index $2.5$ to represent $10\%$ of the $3-20\,$keV flux (paper~III). Although a $10\%$ \HT was ideal in outburst \four, a different level could in principle be necessary in the other three\ outbursts, and the hard tail proxy might vary with time. Because of this proxy, the transition from soft-intermediate to soft states is not clear-cut: some of the states classified as soft could be soft-intermediate, and vice versa. While timing properties are often used to disentangle these two states, we wish to only use the X-ray continuum. As we can see in the large statistical error bars in the fits from \citet{Clavel16}, this is not a major issue for the minimizing procedure since both these states are disk-dominated and $\Gamma$ is often unreliable. However, an important aspect of our model is its ability to link radio and X-ray fluxes, a unique characteristic of disk-driven ejections \citep[][and references therein]{BP82, Ferreira97}. Minor state differences between the soft and soft-intermediate states could then translate into major dynamical differences; see, for example, the soft-intermediate phase of outburst \three where radio flux is predicted by the model but was not observed. It is also possible that the jet is not fully self-similar for small values of transition radius $r_J \gtrsim r_{isco}$ (\ie , during spectrally soft states; paper~IV). In other words, even if a JED is present in the inner regions during the soft-intermediate states, and thus ejections are produced, we do not expect the jet production to be correctly reproduced by our approach. For this reason, although emission is predicted but not observed, we do not believe this represents a fundamental issue with the model.

\subsection{Timing properties} \label{sec:resQPOs}


The \textit{RXTE}/PCA data were already treated in the past in (at least) four comprehensive studies: \citet{Motta11}, \citet{Nandi12}, \citet{Gao14}, and \citet{2017ApJ...845..143Z}. Since we found discrepancies among previous QPO (non-)detections (see, \eg, obsID 95409-01-17-02), we decided to perform our own analysis to get a consistent data set. In order to address this issue in a model-independent way, the QPO types are solely defined using the PDS fitting results of the following procedure (\ie , no time lags or spectral states).



To analyze the timing, we consider GoodXenon, Event, and Binned data modes from all \textit{RXTE}/PCA archival observations of \gx. We reduce the data using the version 6.24 of the HEASOFT. We follow standard procedures\footnote{see https://heasarc.gsfc.nasa.gov/docs/xte/abc/contents.html or https://heasarc.gsfc.nasa.gov/docs/xte/recipes/cook\_book.html} to filter bad time intervals, and extract light curves with bin sizes of $2^{-10}\,$s and $2^{-7}\,$s in three different energy bands to probe different emission regions: spectral channels $6-13$, $14-47$, and $6-47$ (energy bands $2.87-6.12\,$keV, $6.12-19.78\,$keV, and $2.87-19.78\,$keV, PCA calibration epoch 5). We then extract power spectra from the three light curves with the \texttt{powspec} tool, and convert them into XSPEC readable files to ease the fits. We use version 12.10.0c of XSPEC for the PDS fittings.
We use a semiautomatic iterative process to fit the PDS with pyXSPEC. To represent the white noise, we first fit the high-frequency part of the $2^{-10}\,$s bin-sized PDS with a constant (\ie , a power-law slope of $0$). The high-frequency part corresponds to $80-512\,$Hz. Leaving the normalization free to vary allows us to precisely estimate the dead time affected level of white noise \citep[see for example][]{2018ApJ...865..113V}. For the power spectra extracted with a $2^{-7}$ bin size, the maximum frequency is $64\,$Hz, thus we fit the white noise with a frozen flat power law of normalization 2.
We then freeze the parameters of this component and consider the PDS over the entire frequency range. If the fit is valid (\ie , $\chi^2_{\mathrm{tot}, \mathrm{red}} < 1.2$), no additional component is needed and we do not consider the observation further. If the PDS diverges from pure white noise, we first add a zero-centered Lorentzian initialized with all its parameters left free. If the fit statistic is poor ($\chi^2_{\mathrm{tot}, \mathrm{red}} > 1.2$) after the initial fit, we iterate the process by adding a new Lorentzian at the frequency of the largest residuals. This iterative process is repeated up to four times, that is, five total Lorentzians (broad or narrow). When a good fit is achieved, we calculate the coherence factor $Q= f/\Delta f$ of each Lorentzian, where f is the centroid frequency and $\Delta f$ the width. We identify a Lorentzian as a QPO if its coherence factor is $Q > 2$ with a significance $> 5 \sigma$. We note that five Lorentzians are sufficient for all the fits to converge in this paper, and among all the fitted Lorentzians, about $2\%$ of these are detected as QPOs. The QPO type is only derived from its coherence factor (A or B/C) and the presence of source broadband noise (B or C).


\begin{table}[h!]
\centering
\caption{Number of reported (fundamental) LFQPOs in \gx using the \textit{RXTE}/PCA observations. See introduction for the definition of \tA, \tB, and \tC.}
\begin{tabular}{l | c c c c}
\hline & \tA & \tB & \tC \\ \hline
\citet{Motta11} & 7 & 34 & 75 \\ 
\citet{Nandi12} & - & 17 & 22 \\ 
\citet{Gao14} & - & 34 & - \\ 
\citet{2017ApJ...845..143Z} & - & - & 23 \\ \hline
 new QPOs & \newA & \newB & \newC \\ \hline \hline
Total unique QPOs & 7 & 41 & 92 \\ \hline
\end{tabular}
\label{table:QPOs} 
\end{table}

We apply this method to all \textit{RXTE}/PCA observations of \gx , and find a total of $84$, $107$, and $93$ QPOs in the $2.87-6.12\,$keV, $6.12-19.78\,$keV, and $2.87-19.78\,$keV band, respectively. We find no significant differences in the QPO frequency\footnote{For other differences in the continuum, rms, QPO types, or broadband noise, see Marcel et al., in prep.} between the different energy channels and present the results in the largest energy band ($2.78-19.78\,$keV). We summarize these numbers and those obtained in other works using the same X-ray data in Table~\ref{table:QPOs},
. We note that \citet{Motta11} and \citet{Nandi12} searched for LFQPOs in the $2-15\,$keV range, \citet{Gao14} in the $2-24\,$keV range, and \citet{2017ApJ...845..143Z} in the $2-60\,$keV range, explaining the potential differences (see Appendix~\ref{sec:FalseQPOs}). The entire list of the detected LFQPOs using these data can be found in Table~\ref{tab:allqpos}, and new LFQPOs are listed in Table~\ref{tab:newqpos}.

\begin{figure}[h!]
 \includegraphics[width=1.0\linewidth]{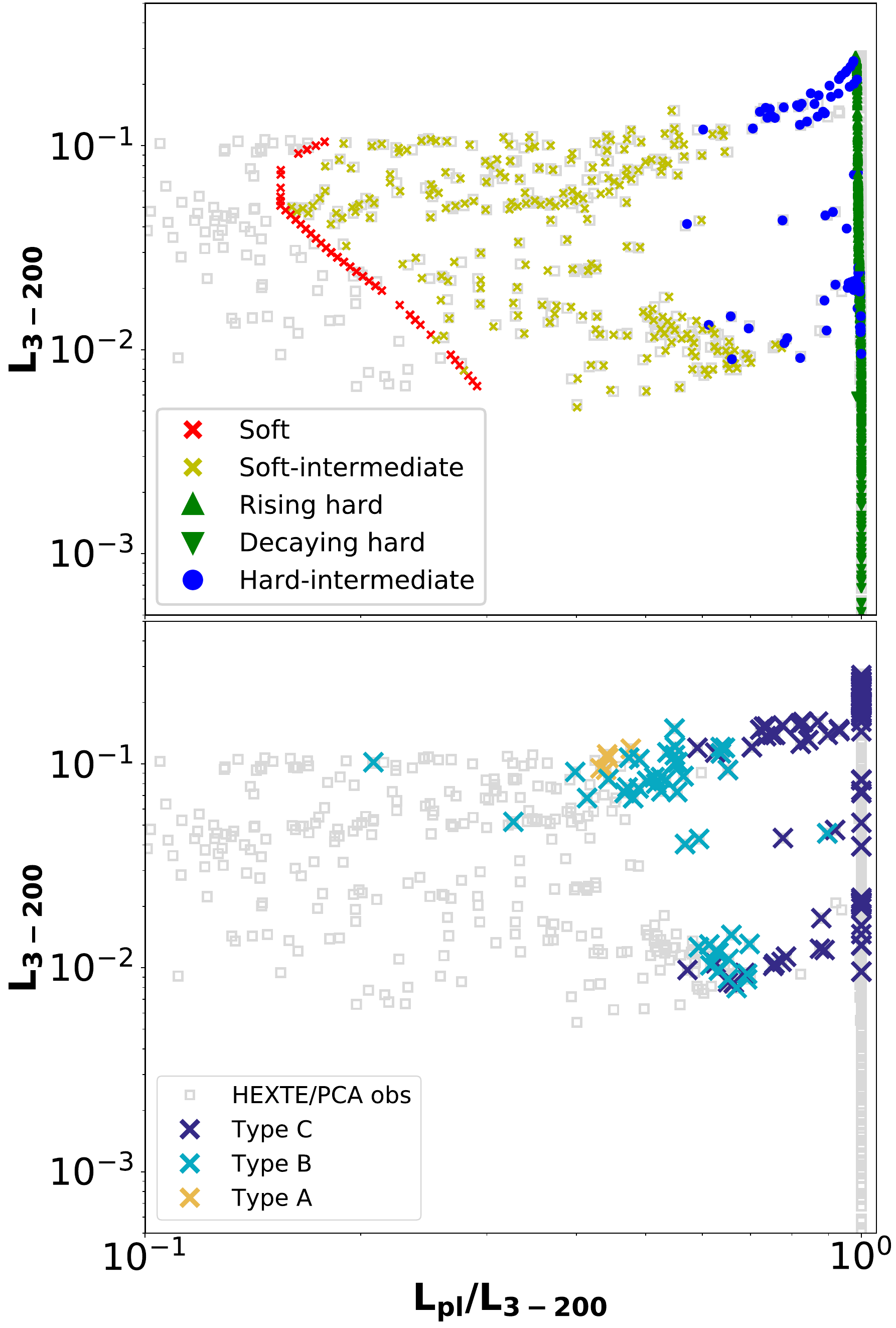}
 \caption{Plots of DFLDs showing the distribution of states (top) and observed LFQPOs (bottom). In the top panel, states are shown in the same color code as in Figs.~\ref{fig:OB12} and \ref{fig:OB34} (see legend). In the bottom panel, the position where QPOs were identified are shown in different colors (see legend). The light gray squares in background represent observations from \citet{Clavel16}.}
 \label{fig:DFLD}
\end{figure}

We plot the observed LFQPOs in the disk fraction luminosity diagrams (DFLDs) at the bottom of Fig.~\ref{fig:DFLD}. As expected, \tC QPOs are detected in the hard state and during the beginning of transitions: \tB mainly during transitions and \tA particularly at the end of the hard-to-soft transition. While this is usually a by-product of the definition of each type, we recall that the QPO types (A, B, or C) and the spectral states (hard, hard-intermediate, soft-intermediate, or soft) are defined using independent methods in this study. We show the time evolution of the QPO frequency in Fig.~\ref{fig:QPOvsTime}. 

\section{LFQPOs and the accretion flow} \label{sec:linkQPOs}

\subsection{LFQPOs and the accretion flow parameters} \label{sec:QpoLink}

In the past, a possible correlation between QPO frequency and the disk flux or accretion rate was observed by multiple authors \citep[see, e.g.,][]{1999AstL...25..718T, 2000MNRAS.312..151R, 2000ApJ...531..537S}. However, such a correlation works only on a short range of QPO frequencies and we focus in this work on the transition radius \rj .

A correlation between the inner radius of the optically thick disk (or truncation radius) and the QPO frequencies was evaluated for different objects in the past (see Sect.~\ref{sec:Correlation}). However, a large fraction of the detected LFQPOs lie at a fairly high luminosity: $73$ out of $140$ at X-ray fluxes above $10\% \, L_{Edd}$. The detection of so many LFQPOs above this luminosity is inconsistent with their production at the optically thick-thin transition. Indeed, at such high luminosity, accretion flow solutions are expected to become optically thick \citep[$\tau \gtrsim 1$;][]{1998MNRAS.301..435Z, 1999ASPC..161..295B} down to the innermost stable circular orbit (ISCO): There is no optically thick-thin transition where the LFQPO can be produced. 
In a JED-SAD configuration, sharp transitions in the structure (\eg , density and temperature) are necessarily observed at the interface between the two flows, (\ie , \rj ). This interface is a natural place for instabilities to develop and is a requirement in most models for the production of QPOs \citep[][]{2020arXiv200108758I}. We thus expect this radius to have an impact on the production of QPOs, and investigate this possibility here. 
The Kepler orbital frequency can be written as $\nu_{K} (r_{J}) = \Omega_K (r_{J}) / (2\pi) = 55.4 \times 10^2 \cdot r_{J}^{-1.5} \,$Hz, where $r_{J}$ is the transition radius in units of $R_g$, and $\Omega_K$ its associated Kepler angular velocity (for a $m=5.8$ black hole).

\begin{figure*}[h!]
 \includegraphics[width=1.0\linewidth]{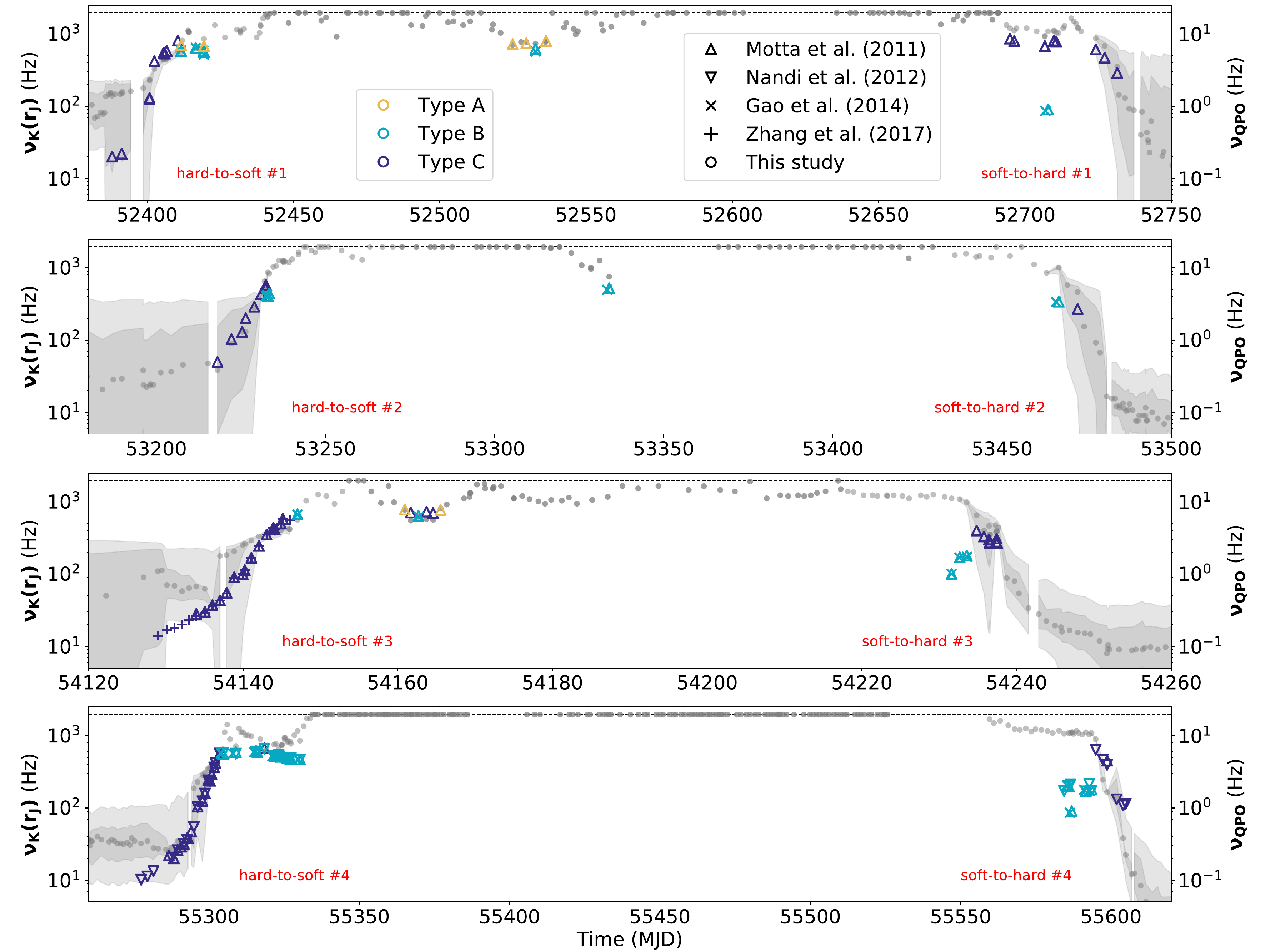} 
 \caption{Co-evolution of the observed QPO frequency $\nu_{QPO}$ and the Kepler rotation frequency $\nu_{K}(r_J)$ for the 4 different outbursts, showing both hard-to-soft and soft-to-hard transitions (see red annotations). The left y-axis shows the Kepler frequency at the transition radius, $\nu_K(r_J)$, and the $5\%$ and $10\%$ confidence intervals. The right Y-axis shows the observed QPO frequency, $\nu_{QPO}$. Different markers are used to distinguish the different works: \citet{Motta11} is indicated in upward triangles, \citet{Nandi12} is indicated in downward triangles, \citet{Gao14} is indicated in crosses, \citet{2017ApJ...845..143Z} is shown in plus signs, and new detections are shown in circles (see text). Different colors indicate the different types of QPOs (see legend). }
 \label{fig:QPOvsTime}
\end{figure*}

A common behavior is observed during transient XrB outbursts: the frequency of the QPO increases in the hard-to-soft transition from $\sim 0.1$ to $\sim 10\,$Hz, and decreases in the soft-to-hard transition from $\simeq 10$ down to $\simeq 0.1\,$Hz. Similarly, the Kepler frequency in \rj increases during the hard-to-soft transition from $\nu_K (r_J = 50-100) \approx 10\,$Hz to $\nu_K (r_J = 2-4) \approx 10^3\,$Hz, and decreases during the soft-to-hard transition from $\sim 10^3\,$Hz to $\sim 10\,$Hz \citep{Remillard06}. This frequency is thus typically two orders of magnitude higher than LFQPO frequencies\footnote{Equivalently, we can link the QPO frequency to a given radius $r_{QPO}$. Such a radius would thus be a factor $\sim 100^{2/3} \sim 20$ bigger than our typical transition radius \rj.} observed.


\subsection{Co-evolution : $\nu_{QPO}(t)$ and $\nu_{K}(r_J)(t)$}

We show in Fig.~\ref{fig:QPOvsTime} the Kepler frequency $\nu_K (r_J) = \Omega_K (r_J) / (2 \pi )$ of our four outbursts in gray circles, from top to bottom: \one, \two, \three, and \four. The $5\%$ and $10\%$ confidence regions are also shown, but the color associated with each state is removed for clarity. Additionally, we overplot the frequency of all observed LFQPOs from this study and published data (Sect.~\ref{sec:resQPOs}). The two y-axes, $\nu_K (r_J)$ on the left and $\nu_{QPO}$ on the right, are shifted with respect to each other by two orders of magnitude: $\nu_{QPO} = \nu_K (r_J) / 100$. The global trend of $\nu_K (r_J)$ follows closely that observed for \tA and \tC LFQPOs, especially the four hard-to-soft transitions (on the left side of each panel).
The JED-SAD model is a theoretical model built to explain the possible dynamical structure of the accretion flow, that is able to reproduce the X-ray and radio global spectral shapes in outbursts. Our minimization procedure does not directly include the temperature (or flux) of the soft component, which are the parameter(s) usually used to derive the truncation radius and compare to QPO frequencies. For these two reasons, the apparent match shown in Fig.~\ref{fig:QPOvsTime} had not been anticipated.

It is also remarkable that there are no QPOs detected when $r_J = r_{isco}$, that is, when $\nu_K (r_J)$ follows the horizontal black dotted line. In other words, there are no QPOs detected when only a SAD is needed to reproduce the X-ray spectra. This result is consistent with a production of QPOs related to the inner hot accretion flow and also justifies the distinction\footnote{We recall here that the spectral states were defined solely on the X-ray continuum (disk and power-law), independently from the presence of any (type of) QPO.} between soft and soft-intermediate states (Sect.~\ref{sec:res}).
There are however a few ill-behaved regions where $\nu_{QPO}$ does not follow $\nu_K (r_J)$, especially when \tB are detected; see for example the soft-to-hard transitions of outbursts \one and \four.


\subsection{Correlation with Kepler frequency} \label{sec:Correlation}

\begin{figure}[h!]
 \includegraphics[width=1.0\linewidth]{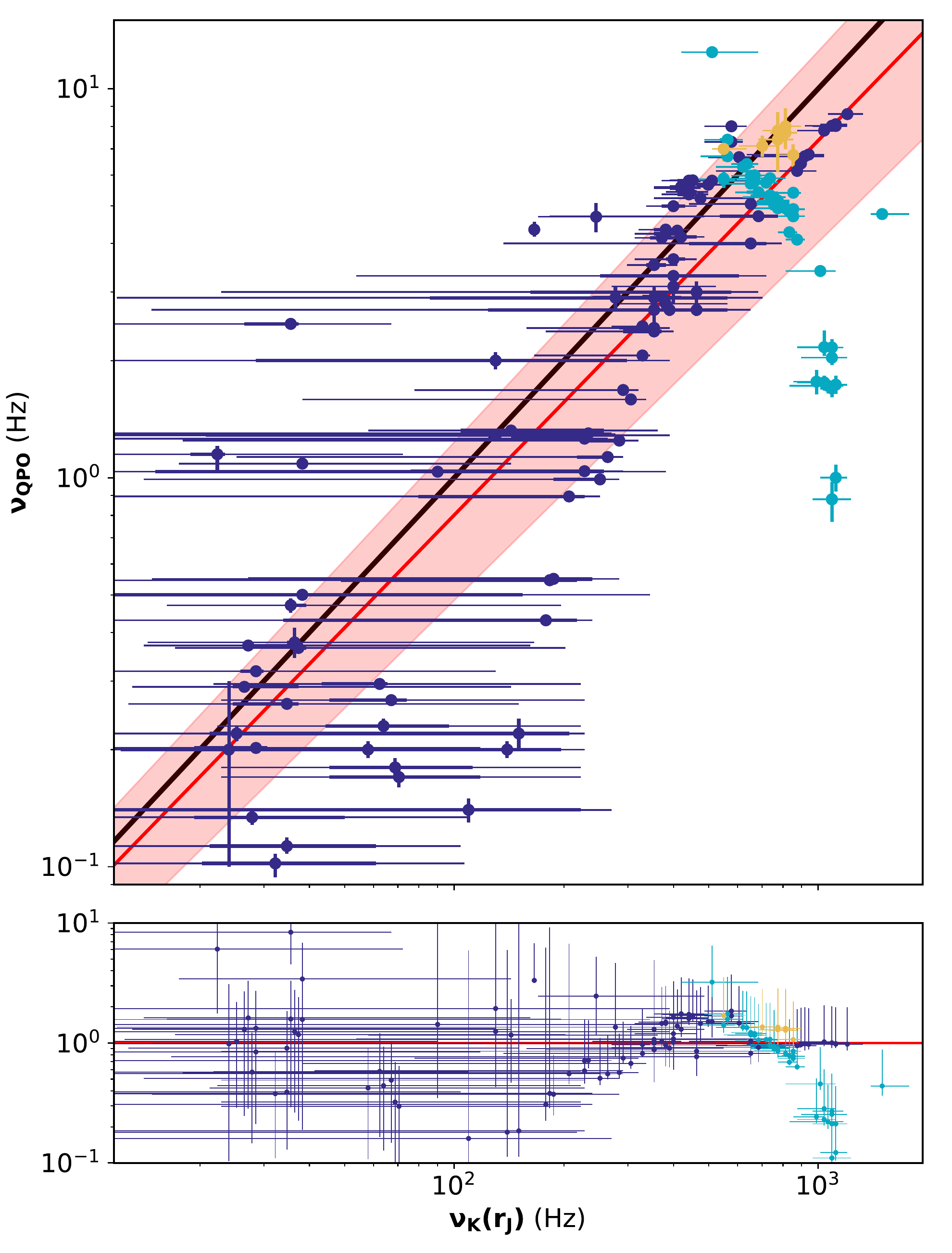}
 \caption{Correlation between observed frequency $\nu_{QPO}$ and the Kepler frequency at the transition radius $\nu_{K}(r_J)$ (top) and the residuals defined as the ratio of the best weighted fit (in red) to the data (bottom). \tA QPOs are indicated in light yellow, \tB in cyan, and \tC in dark blue. The red line indicates the weighted linear fit considering only \tC LFQPOs, $\nu_{QPO} = (10 \pm 3) \times 10^{-3} \cdot \nu_{K} (r_J)^{0.96 \pm 0.04}$; the black line indicates the $\nu_{QPO} = \nu_K (r_J) / 100$ line.}
 \label{fig:QPOvsKepler}
\end{figure}

To ascertain the possibility of a correlation between the accretion flow and these LFQPOs, we show $\nu_{QPO}$ as function of $\nu_{K} (r_J)$ in Fig.~\ref{fig:QPOvsKepler}. 
We use a weighted linear regression to investigate a possible correlation with \tC QPOs, and find a best fit
\begin{equation}
\nu_{QPO} = (10 \pm 3) \times 10^{-3} \cdot \nu_{K} (r_J)^{0.96 \pm 0.04} \nonumber.
\end{equation}
\noindent This correlation is drawn in red on Fig.~\ref{fig:QPOvsKepler}. This is an excellent correlation considering the important error bars implied and the simplicity of the fitting procedure. Notably, the index $0.96 \pm 0.04$ ($1 \sigma$ error), close to $1$, suggests that $\nu_{QPO} \propto \nu_K (r_J)$. When forcing a slope of $1$, the best weighted fit results in $\nu_{QPO} \simeq (7.5 \pm 0.2) \times 10^{-3} \cdot \nu_{K} (r_J)$, that is, a factor $q = 133 \pm 4$ between $\nu_{QPO}$ and $\nu_K (r_J)$. 

This is the first time that such a correlation is found for \gx , but multiple studies found very similar results for other XrBs: GRS 1915+105 \citep{Muno99, 1999AstL...25..718T, Rodriguez02}, XTE J1748-288 \citep{2000MNRAS.312..151R}, XTE J1550-564 \citep{2000ApJ...531..537S, 2004ApJ...612.1018R}, and GRO J1655-40 \citep{2000ApJ...531..537S}. In these studies, a correlation between the frequency of the \tC QPOs and the inner radius of an optically thick disk (\ie , similar to \rj here at low luminosities), and showed correlations $\nu_{QPO} \sim \nu_K (r_J) / q$, where $q$ is in the range $50-150$. In our study, we found $q=133 \pm 4$, and varying between outbursts, $q_{\# 1} = 133 \pm 5$, $q_{\# 2} = 97 \pm 5$, $q_{\# 3} = 136 \pm 5$, and $q_{\# 4} = 75 \pm 11$ (see from Figs.~\ref{fig:QPOvsKepler1} to \ref{fig:QPOvsKepler4}). A variation in $q$ between outbursts or objects would be a major constraint for QPO models, but unfortunately the lack of statistics and the important error bars prevent us from drawing any conclusion with the present study.

Additionally, although we only fitted using \tC LFQPOs, the best fit also seems to capture \tA . However, \tB QPOs do not follow the same correlation. \tB have a much lower frequency than the correlation in Fig.~\ref{fig:QPOvsKepler}, by a factor of up to $\sim 10$. The factor $q \sim 100$ between $\nu_{QPO}$ and $\nu_K (r_J)$ now needs to be as high as $\sim 1000$. It is hard to imagine a process, even secular, to resolve this discrepancy. For this reason, it would be natural to consider that \tB are produced via a process different from \tA and \tC. However, \tB are detected at tiny transition radii (\ie , high Kepler frequencies), when relativistic effects are important. Rather than the usual Kepler frequency, we examine below whether\ using the epicyclic frequency at the transition radius could solve this issue \citep[see][]{Varniere02}.

\subsection{Correlation with transition radius}

\begin{figure}[h!]
 \includegraphics[width=1.0\linewidth]{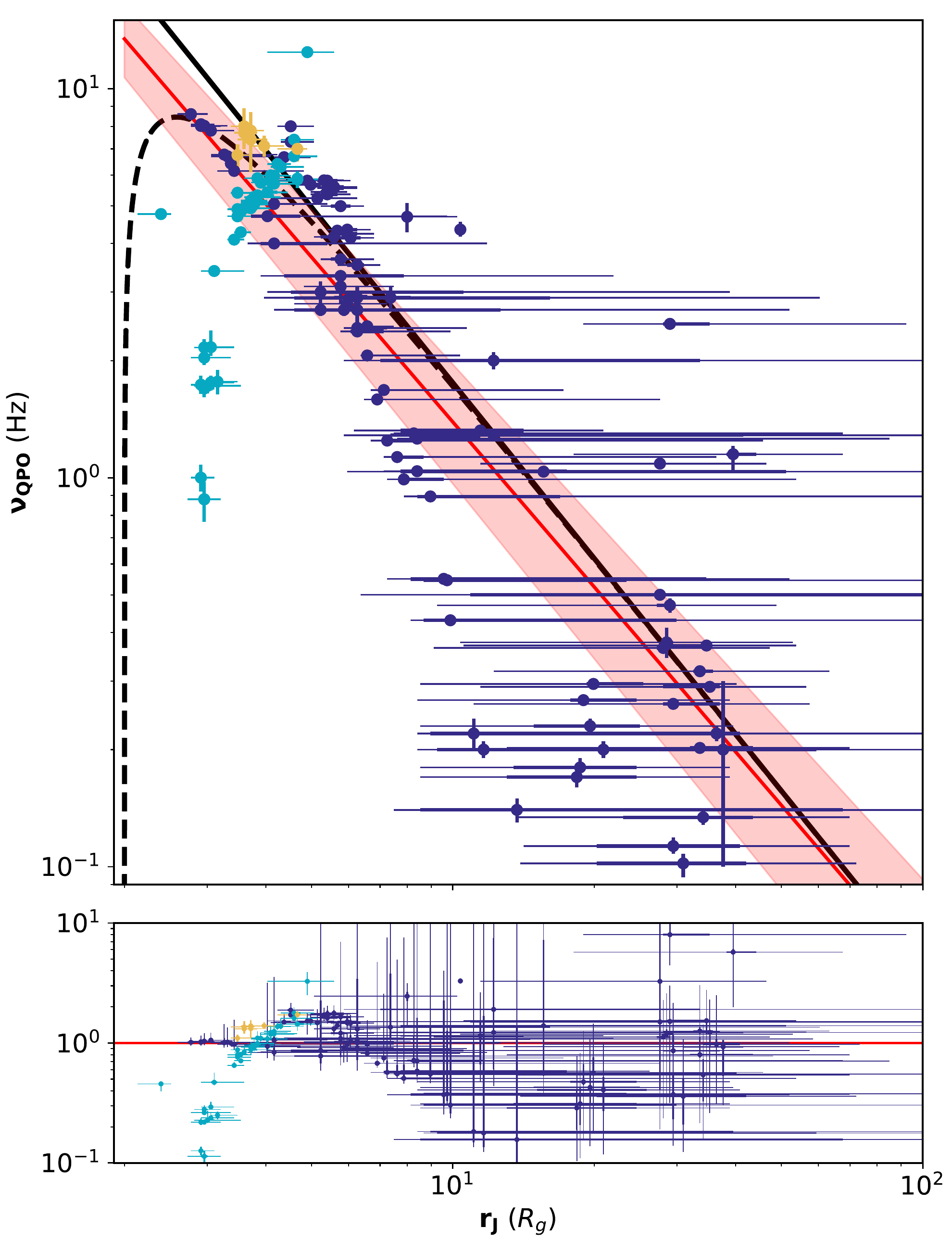}
 \caption{Correlation between observed frequency $\nu_{QPO}$ and the transition radius $r_J$ (top) and the residuals defined as the ratio of the best weighted fit (red) to the data (bottom). \tA QPOs are shown in light yellow, \tB are indicated in cyan, \tC are shown in dark blue. The weighted linear fit considering only \tC LFQPOs, $\nu_{QPO} = (36 \pm 6) \cdot r_J^{-1.41 \pm 0.08}$, is indicated as a red line, the solid black line represents $\nu_{QPO} = \nu_K (r_J) / 100$, and the dashed black line represents $\nu_{QPO} = \nu_{\mathrm{ep}} (r_J) / 100$.}
 \label{fig:QPOvsRj}
\end{figure}

We show the correlation between $\nu_{QPO}$ and \rj in Fig.~\ref{fig:QPOvsRj}. Owing to the effects of general relativity, in this section we should consider the epicyclic frequency \citep{Varniere02},
\begin{equation}
\nu_{\mathrm{ep}} = \nu_K (r_{J}) \cdot \left( 1 - \frac{r_{isco}^2}{r_J^2} \right) = 5.54 \times 10^3 \cdot \left( 1 - r_{isco}^2 r_J^{-2} \right) r_{J}^{-3/2} \, \mathrm{Hz} \label{eq:kappaQPO}
,\end{equation}
\noindent where we retrieve $\nu_{\mathrm{ep}} = \nu_K (r_J)$ for $r_J \gg r_{isco}$. Because these two frequencies are similar until $r_J \gtrsim r_{isco}$, $\nu_K (r_J)$ and $\nu_{\mathrm{ep}} (r_J)$ follow a similar correlation with frequencies two orders of magnitude too high $\nu_{QPO} = \nu_{\mathrm{ep}} (r_J) / 100$. Both these correlations are shown in Fig.~\ref{fig:QPOvsRj}: in solid black $\nu_{QPO} = \nu_K (r_J) / 100$ and in dashed black $\nu_{QPO} = \nu_{\mathrm{ep}} (r_J) / 100$. Again, the black curve aligns with the best fit (red line) well.

Considering this new $\nu_{QPO} = \nu_{\mathrm{ep}} (r_J) / 100$ slope, what resembles a shift by a factor $10$ in $\nu_{QPO}$ in Fig.~\ref{fig:QPOvsKepler} can be interpreted as a shift by a factor $\sim 1.5$ in \rj in Fig.~\ref{fig:QPOvsRj}.
If \rj was smaller than the value calculated by our model, say $r_J = 2.1-2.5$ instead of $r_J \sim 3-4$, then these \tB QPOs could in principle fit in the dashed-black correlation\footnote{While we argued earlier that the self-similar assumption might not hold for the smallest values of \rj , this argument is irrelevant regarding QPOs. The possibility of having QPOs even for $r_J \gtrsim r_{isco}$ depends on the processus involved in their production.}.
An overestimate of \rj for \tB QPOs can have two major causes. First, our model does not include all possible effects; a different transition radius could be obtained if we include a relativistic treatment of the equations, ray-tracing effects, or gravitational redshifts for example. Second, as discussed previously in Sect.~\ref{sec:res}, the difference between soft-intermediate ($r_J \sim 3-4$) and soft states ($r_J = r_{isco} = 2$) is not clear cut in our fits. A better treatment of the \HT could very well lead to over- or underestimates of the transition radius once such small values are reached.
%
However, we note that \tC QPOs are well behaved for even the smallest transition radii ($r_J < 3$), while \tB QPOs have already diverged from the correlation at $r_J = 4-5$. This difference suggests that small transition radii are not primarily responsible for pushing \tB QPOs off the correlation above. Although it is tempting to try to correct the transition radius, we believe this is premature given the points above. Instead, these features likely follow a different correlation entirely, as some previous studies have suggested already \citep[see, \eg,][]{Motta11}. We would like to point out that the transition from some \tC to \tB appears smooth in the two residuals at the bottom of Figs.~\ref{fig:QPOvsKepler} and \ref{fig:QPOvsRj}, suggesting further relationships between \tB and \tC QPOs that will be explored in a future work.

\section{Summary and conclusions} \label{sec:Ccl}

In this paper, we assume that the accretion flow is separated into two regions: an inner JED and an outer SAD. This is known as the JED-SAD paradigm. We address some timing properties of the hybrid JED-SAD configuration for the first time, using tools and methods detailed in the previous papers of this series. We summarize the methods and major results from this study in three points:

First, we select \textit{RXTE}/PCA (X-ray) and ATCA (radio) data covering three outbursts of \gx. We then use the procedure, detailed in paper~IV for a previous outburst, to obtain an estimate of \pair in the JED-SAD paradigm during these three new outbursts. The dynamical insights from the study of the four outbursts will be discussed in a forthcoming paper, but it is already important to note that a unique multiplication factor $\tilde{f}_R$ was used to estimate the radio flux. In other words, the jet radiative efficiency is constant during all four outbursts, although the jets are quenched during each soft state.

Second, we use the same \textit{RXTE}/PCA data and focus on a study of LFQPOs. We report the detection of seven new LFQPOs (see Table~\ref{tab:newqpos}). 

Third and more importantly, we confirm that the frequency of \tC QPOs can be linked to the transition radius between two different accretion flows (JED and SAD in this work). This correlation requires a (surprising) factor $q$ between the Kepler frequencies at the transition radius and the QPO frequencies, varying between $\sim 70$ and $\sim 140$ for different outbursts, whose value $q = 133 \pm 4$ for all four outbursts combined. Such a factor is hard to estimate, but multiple studies have been shown to require two orders of magnitude between the QPO and Kepler frequencies around different objects (see Sect.~\ref{sec:linkQPOs}). However, while these former studies envisioned that QPOs were linked to the optically thick-to-thin transition in the flow, we argue for and show a link between QPOs and the transition radius between the inner magnetized ($\mu \sim 0.5$) and an outer weakly magnetized ($\mu \ll 1$) flow. This link indicates that QPOs must be created by a secular (given the high value of $q$) instability or process related to the dynamical JED-SAD transition and thus, somehow, connected to the jets. This connection is consistent with previous suggestions of a link between the existence of jets and the presence of QPOs \citep{Fender09}, but this is, to our knowledge, the first time that the direct link is shown \citep[see section~4.3 in][]{2014SSRv..183..453U}. Moreover, this connection is supported by the observed relations between IR and X-ray QPOs \citep[][]{2010MNRAS.404L..21C, 2016MNRAS.460.3284K, 2019ApJ...887L..19V}, which is a behavior that current models fail to address. 

Because of the remarkable correlation between \tC QPOs and \rj , these results provide independent support to the JED-SAD paradigm. However, the apparent differences between QPO types need to be addressed and the processe(s) to produce the LFQPOs in this paradigm remain to be discussed.

\begin{acknowledgements}
 We thank the anonymous referee for helpful comments and careful reading of the manuscript. The authors acknowledge funding support from the french research national agency (CHAOS project ANR-12-BS05-0009, http://www.chaos-project.fr), \textit{centre national d'Ã©tudes spatiales} (CNES), and the \textit{programme national des hautes Ã©nergies} (PNHE). This research has made use of data, software, and/or web tools obtained from the high energy astrophysics science archive research center (HEASARC), a service of the astrophysics science division at NASA/GSFC. Figures in this paper were produced using the \textsc{matplotlib} package \citep{plt}, and weighted fits using the SciPy package \citep{2019arXiv190710121V}.
\end{acknowledgements}

\bibliographystyle{aa} 
\bibliography{Research.bib}

\clearpage

\appendix

\section{Discrepancies on QPO identifications} \label{sec:FalseQPOs}

In this section, we provide the fitting results of the PDS to illustrate the differences between this work and previous studies.

\begin{figure}[h!]
 \center
 \includegraphics[width=0.9\linewidth]{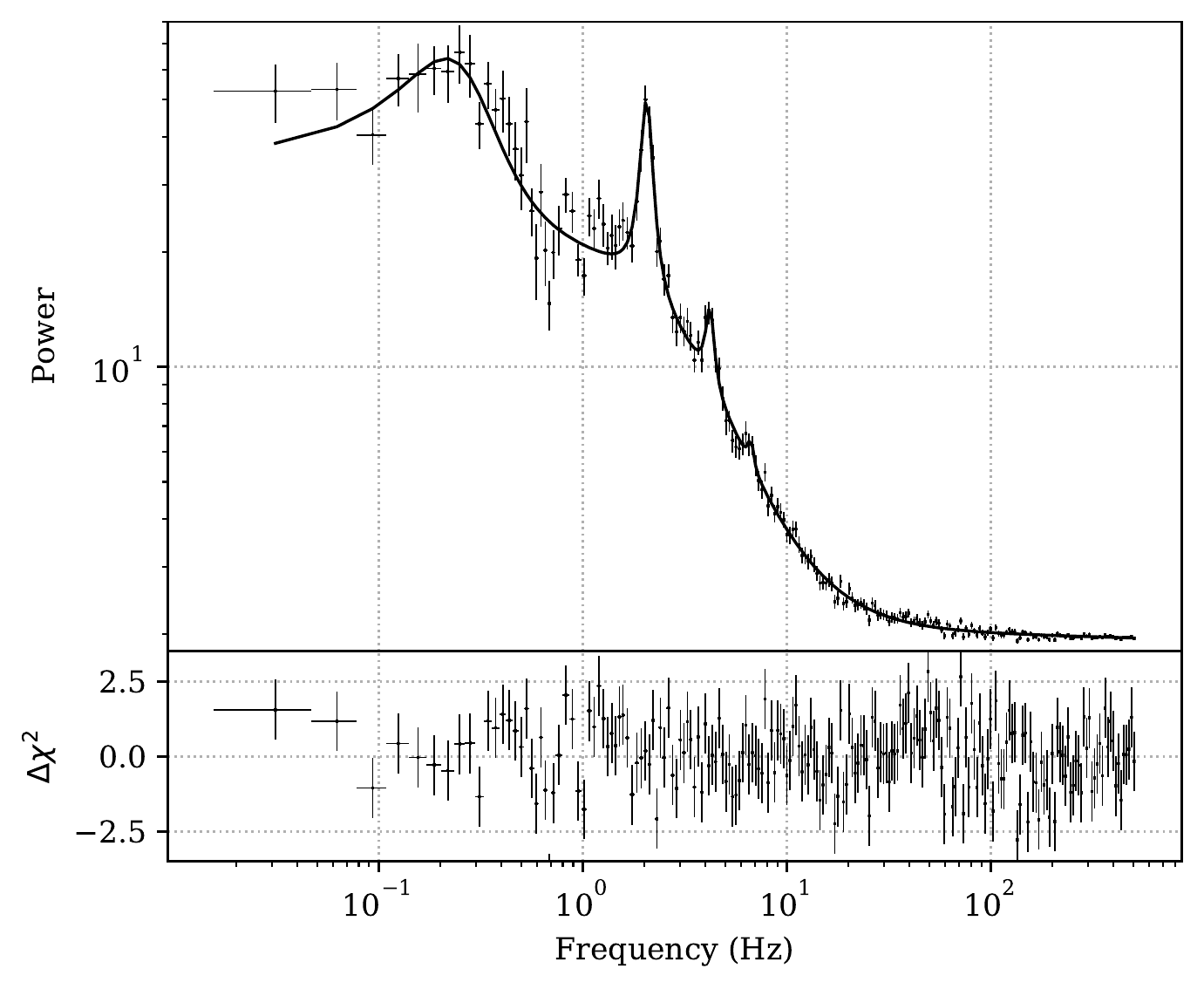}
 \caption{Power spectrum of obsID 70110-01-10-00.}
 \label{fig:70110-01-10-00}
\end{figure}

In obsID 70110-01-10-00, \citet{Motta11} reported a $4.20^{+0.08}_{-0.08}\,$Hz \tC QPO. However, when performing the fit, we believe that this component is anharmonic for two major reasons (see Fig.~\ref{fig:70110-01-10-00}). First, there is a very strong component at $2.06^{+0.01}_{-0.01}\,$Hz with very high coherence factor $Q = 8.16$. Second, there is another harmonic component at $6.67^{+0.24}_{-0.29}\,$Hz, consistent with being three times the frequency of the $2.06^{+0.01}_{-0.01}\,$Hz QPO. We thus have the fundamental at $\nu_0 = 2.06\,$Hz, and two harmonic components at $\nu_1 = 4.20\,\text{Hz} \simeq 2 \nu_0$ and $\nu_2 = 6.67\,\text{Hz} \simeq 3 \nu_0$. 

\begin{figure}[h!]
 \center
 \includegraphics[width=0.9\linewidth]{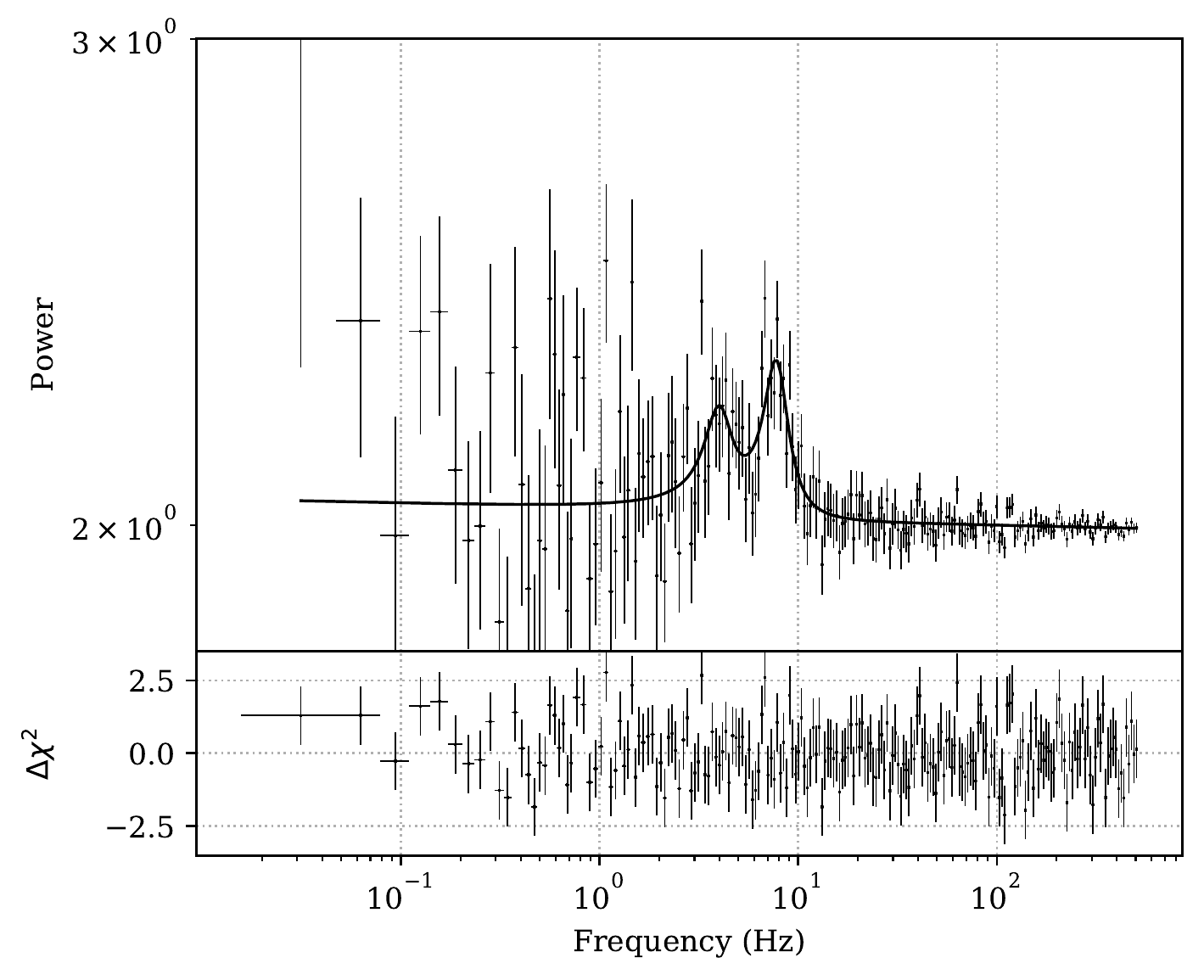}
 \caption{Power spectrum of obsID 92085-01-03-00.}
 \label{fig:92085-01-03-00}
\end{figure}

In obsID 92085-01-03-00, two major components are present at $3.96\,$Hz and $7.72\,$Hz (see Fig.~\ref{fig:92085-01-03-00}). While the components have similar coherence factors $Q = 2.87$ for the $7.72\,$Hz QPO and $Q = 2.52$ for the $3.96\,$Hz QPO, the time evolution is critical. Both QPOs before and after obsID 92085-01-03-00 have frequencies closer to $7.72\,$Hz (see Table~\ref{tab:allqpos}), suggesting that the QPO we are following in this rising outburst is at $7.72\,$Hz. We agree with \citet{Motta11} about the fundamental component. However, they reported this QPO as a \tC, and we strongly believe it should be classified as a \tB for two reasons. First, the previous and following QPOs were \tB (ignoring \tA, see Table~\ref{tab:allqpos}). Second, the power spectrum does not show any broadband noise, a key characteristic of \tC QPOs. We thus decided to keep the QPO of highest frequency, but we classified it as a \tB, unlike \citet{Motta11}.

\begin{figure}[h!]
 \includegraphics[width=0.9\linewidth]{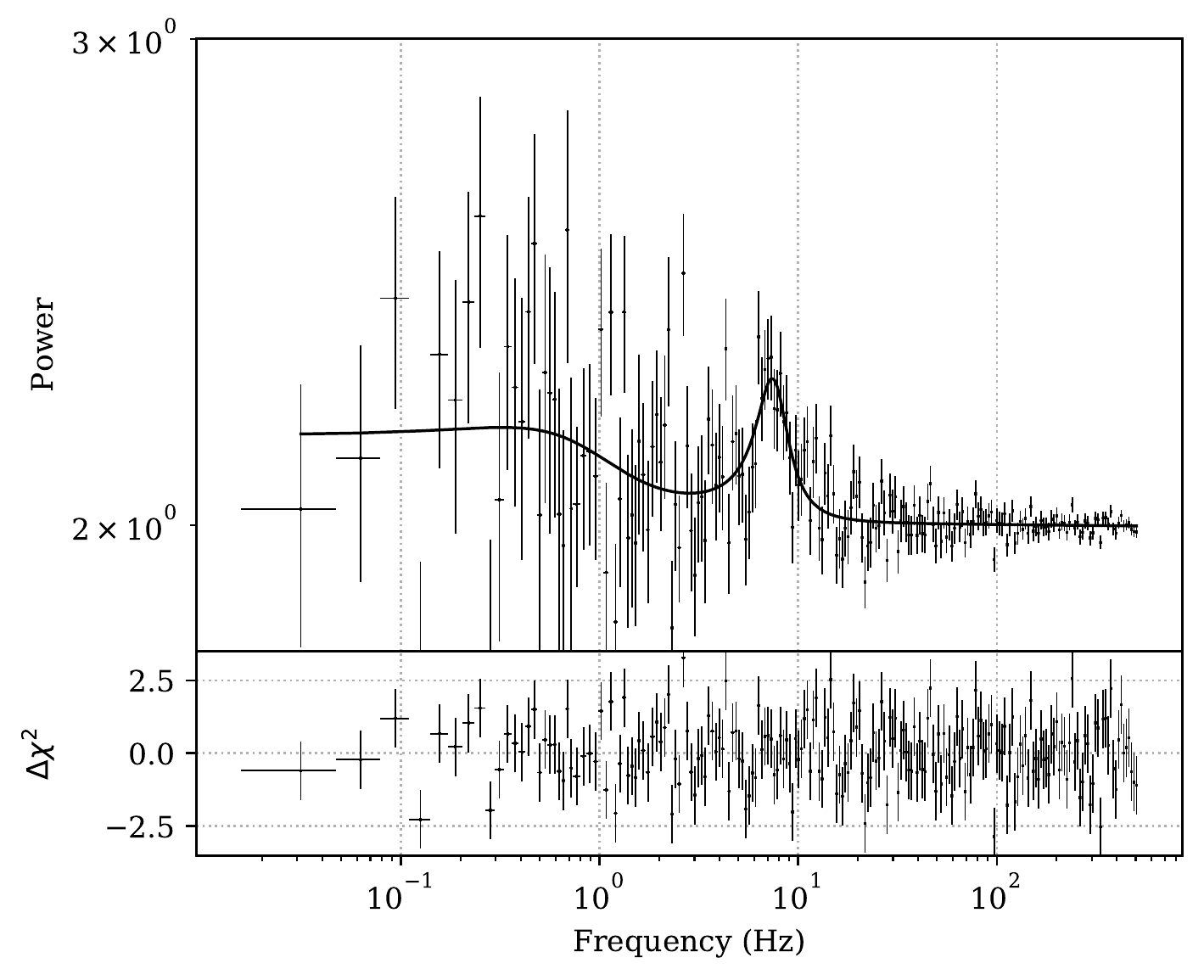}
 \caption{Power spectrum of obsID 92085-01-03-03.}
 \label{fig:92085-01-03-03}
\end{figure}

Similarly, in obsID 92085-01-03-03, \citet{Motta11} report the detection of a \tC QPO at $7.00\,$Hz. However, the power spectrum does not show significant broadband noise (see Fig.~\ref{fig:92085-01-03-03}). This could very well be a result of the different procedure used, or a matter of definition, but to be fully consistent within our study we decided to keep that QPO as a \tB.

\begin{figure}[h!]
 \includegraphics[width=0.9\linewidth]{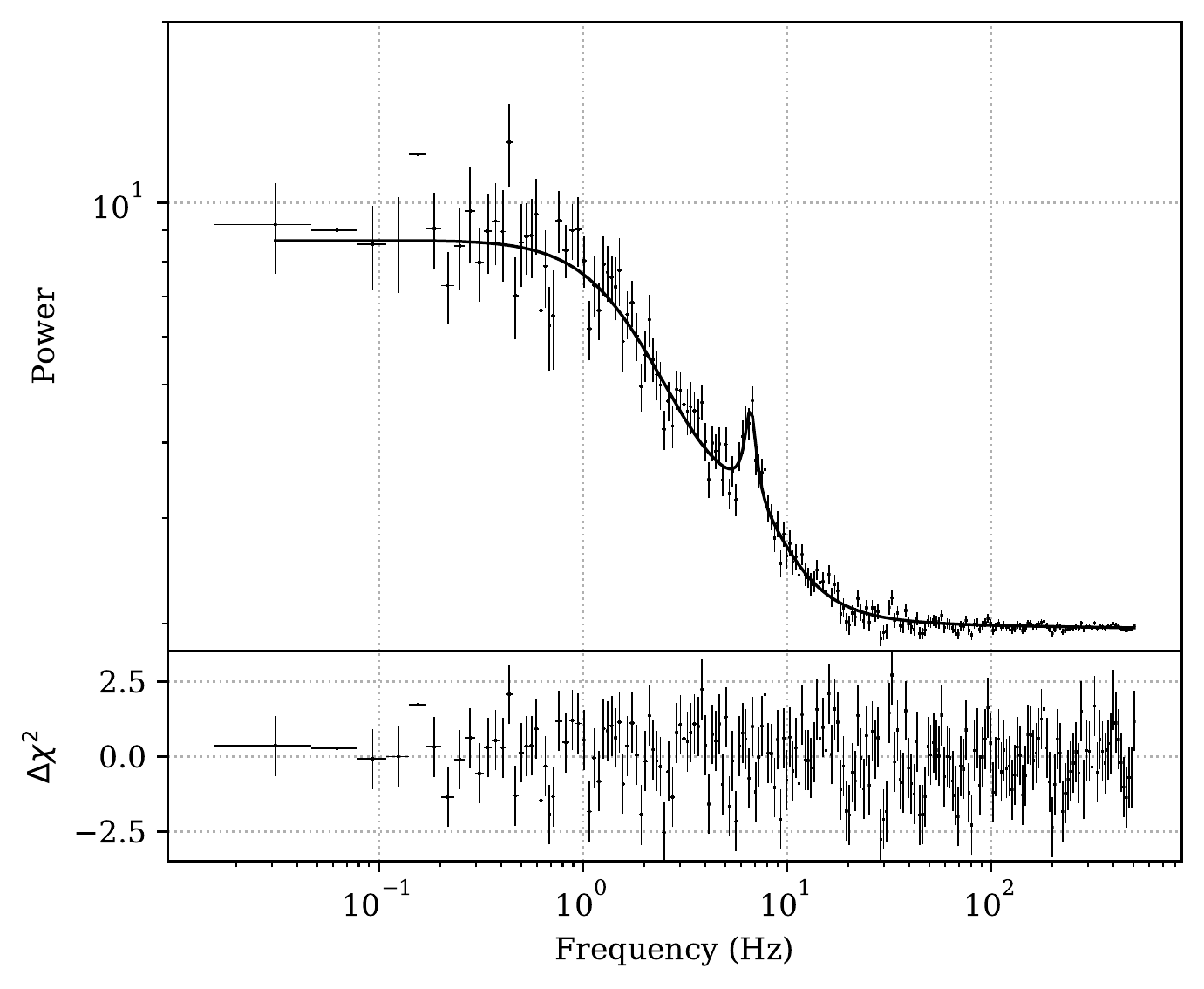}
 \caption{Power spectrum of obsID 95409-01-17-02.}
 \label{fig:95409-01-17-02}
\end{figure}

Finally in obsID 95409-01-17-02, \citet{Motta11} report a $6.67\,$Hz \tC QPO, while \citet{Nandi12} report a $6.65\,$Hz \tB QPO. While the time evolution seems to suggest that this QPO is a \tB (see Table~\ref{tab:allqpos}), we identify this to be a $6.66\,$Hz \tC QPO, because of the presence of a significant broadband noise in the power spectrum (see Fig.~\ref{fig:95409-01-17-02}).

\section{Correlation with Kepler frequency for each outburst} \label{sec:CorrelationsOmegaK}

In this section, we show the correlation of the QPO frequency $\nu_{QPO}$ as function of the Kepler frequency at the transition radius $\nu_{K}(r_J)$ for all four outbursts.

\begin{figure}[h!]
 \includegraphics[width=.9\linewidth]{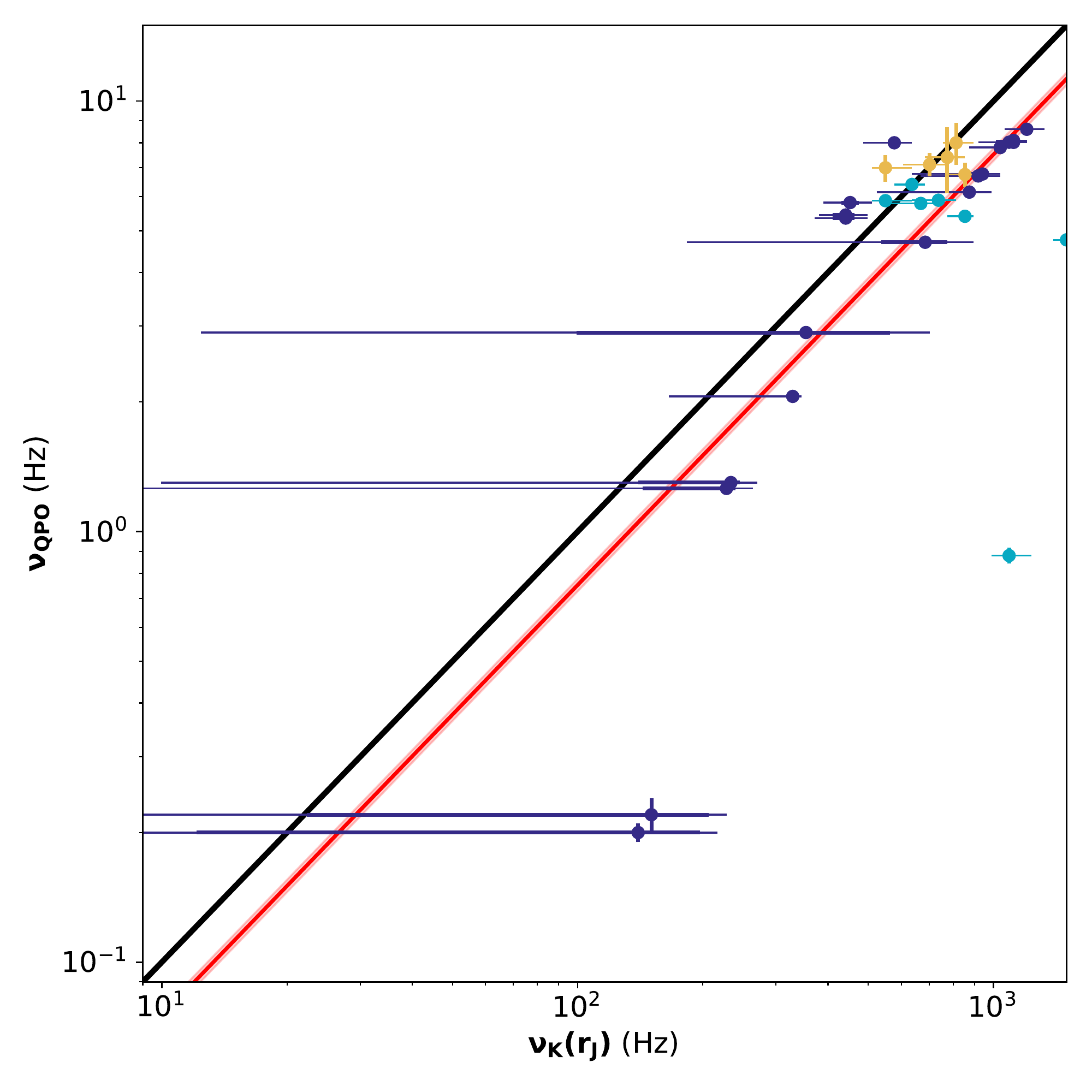}
 \caption{Correlation between observed frequency $\nu_{QPO}$ and the Kepler frequency at the transition radius $\nu_{K}(r_J)$ for outburst \one. Various types of QPOs are shown: \tA are indicated in light yellow, \tB in cyan, and \tC in dark blue. The weighted linear best fit with slope 1 considering only \tC LFQPOs, $\nu_{QPO} = (7.5 \pm 0.3) \times 10^{-3} \cdot \nu_{K} (r_J)$, is indicated in red; the $\nu_{QPO} = \nu_K (r_J) / 100$ line is shown in black.}
 \label{fig:QPOvsKepler1}
\end{figure}

\begin{figure}[h!]
 \includegraphics[width=.9\linewidth]{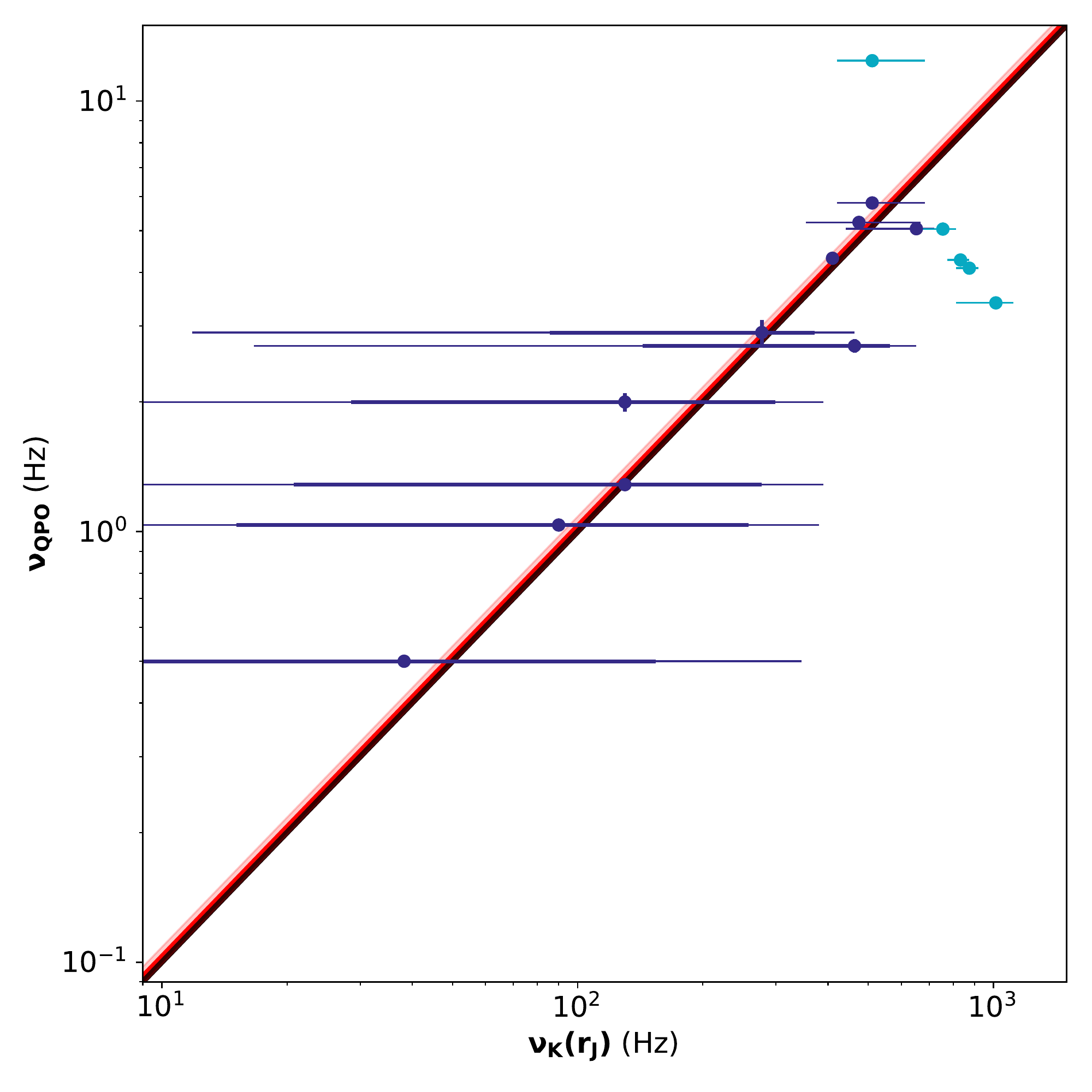}
 \caption{Correlation between observed frequency $\nu_{QPO}$ and the Kepler frequency at the transition radius $\nu_{K}(r_J)$ for outburst \two. This figure is similar to Fig.~\ref{fig:QPOvsKepler1}, but this time $\nu_{QPO} = (10.4 \pm 0.5) \times 10^{-3} \cdot \nu_{K} (r_J)$.}
 \label{fig:QPOvsKepler2}
\end{figure}

\begin{figure}[h!]
 \includegraphics[width=.9\linewidth]{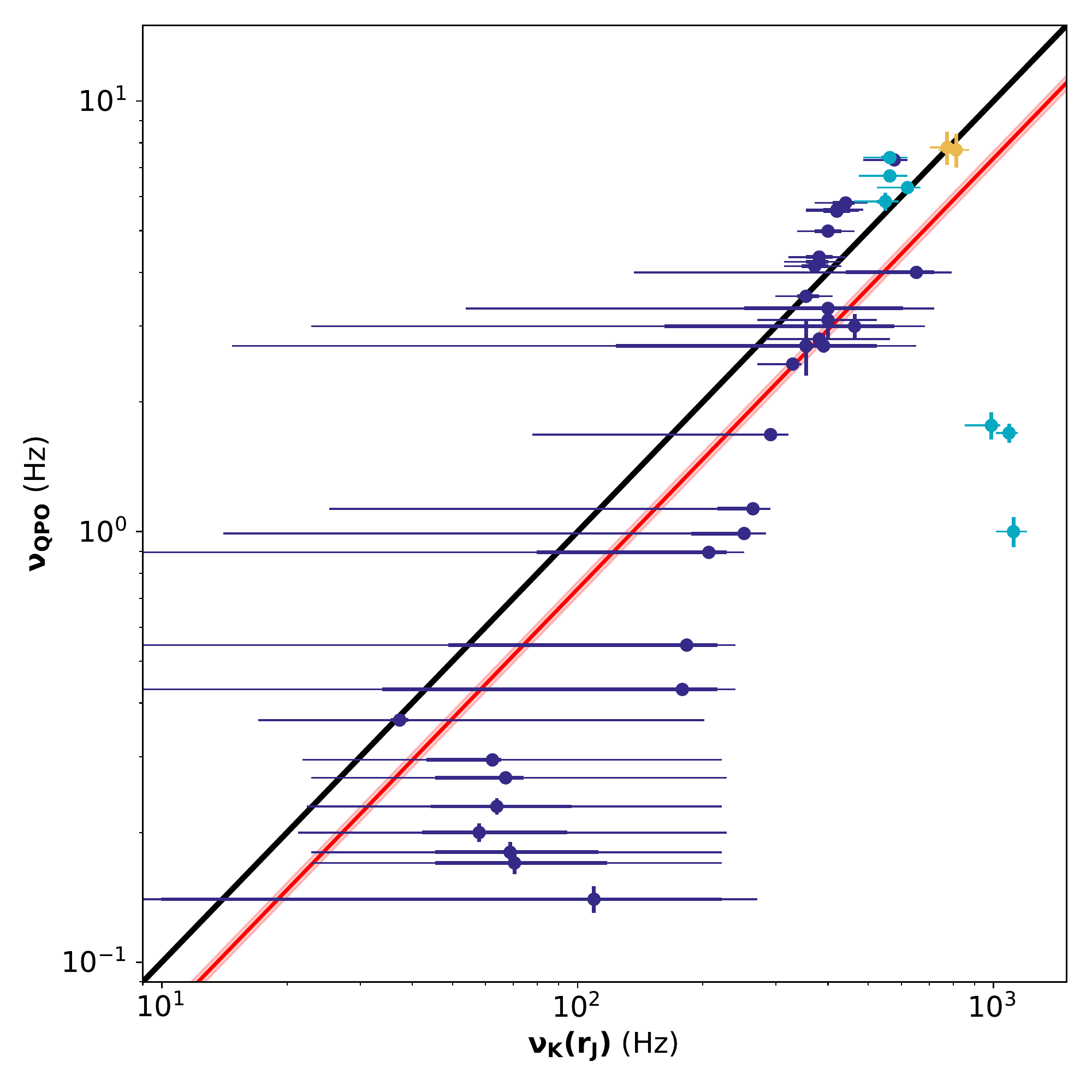}
 \caption{Correlation between observed frequency $\nu_{QPO}$ and the Kepler frequency at the transition radius $\nu_{K}(r_J)$ for outburst \three. This figure is similar to Fig.~\ref{fig:QPOvsKepler1}, but this time $\nu_{QPO} = (7.4 \pm 0.3) \times 10^{-3} \cdot \nu_{K} (r_J)$.}
 \label{fig:QPOvsKepler3}
\end{figure}

\begin{figure}[h!]
 \includegraphics[width=.9\linewidth]{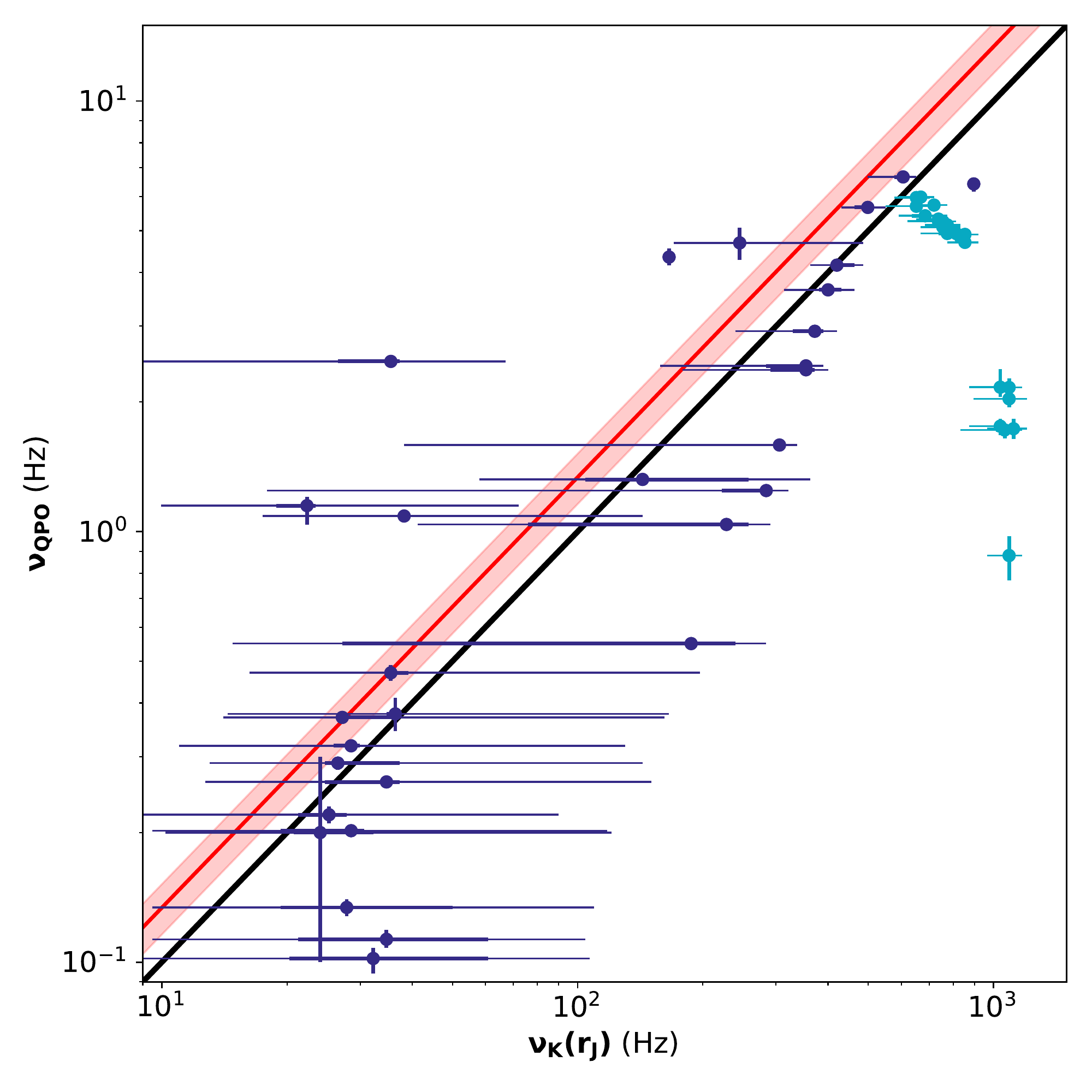}
 \caption{Correlation between observed frequency $\nu_{QPO}$ and the Kepler frequency at the transition radius $\nu_{K}(r_J)$ for outburst \four. This figure is similar to Fig.~\ref{fig:QPOvsKepler1}, but this time $\nu_{QPO} = (13.4 \pm 0.2) \times 10^{-3} \cdot \nu_{K} (r_J)$.}
 \label{fig:QPOvsKepler4}
\end{figure}

\section{Detected LFQPOs} \label{sec:NewQPOs}

We report in the following tables the QPOs detected in this study. See Sect.~\ref{sec:data} for the data involved and Sect.~\ref{sec:resQPOs} for the method used. In the first table we show all the fundamentals detected QPOs in the literature and in this study; in the second table we show only the new QPOs from this study.

\onecolumn 

\newpage

\centering
\tablefirsthead{\toprule ObsID & Date & $\nu_{QPO}$ & Type & Reference & \rj & $\displaystyle \nu_K(r_J)$ & Notes \\ 
& (MJD) & (Hz) & & & $(R_g)$ & (Hz) & \\ \midrule}
\tablehead{%
\multicolumn{8}{l}{Continued from previous page} \\
\toprule
ObsID & Date & $\nu_{QPO}$ & Type & Reference & \rj & $\displaystyle \nu_K(r_J)$ & Notes \\ 
& (MJD) & (Hz) & & & $(R_g)$ & (Hz) & \\ \midrule}
\tabletail{%
\midrule \multicolumn{8}{r}{Continued on next page} \\ \midrule}
\tablelasttail{%
\\\midrule
\multicolumn{8}{r}{{Concluded}} \\ \bottomrule}
\topcaption{Columns 1 to 5: Fundamental quasi-periodic oscillations detected in \textit{RXTE}/PCA data from \gx. Columns 6 and 7: associated transition radius and its Kepler rotation frequency in the JED-SAD paradigm (see paper~IV and section \ref{sec:res}).}
\label{tab:allqpos}
\begin{supertabular}{cccclccc}
40031-03-02-05 & 52388.05 & $0.20^{+0.01}_{-0.01}$ & C & \citet{Motta11} & $11.6^{+232.0}_{-2.7}$ & $140^{+138}_{-67}$ & \\ [3pt] 
\hline
70109-01-05-01 & 52391.32 & $0.22^{+0.02}_{-0.02}$ & C & \citet{Motta11} & $11.1^{+187.6}_{-2.4}$ & $150^{+148}_{-67}$ & \\ [3pt] 
\hline
70109-01-06-00 & 52400.83 & $1.26^{+0.01}_{-0.01}$ & C & \citet{Motta11} & $8.4^{+22.8}_{-0.5}$ & $228^{+196}_{-23}$ & \\ [3pt] 
\hline
70108-03-01-00 & 52400.85 & $1.30^{+0.01}_{-0.01}$ & C & \citet{Motta11} & $8.3^{+19.8}_{-0.5}$ & $234^{+196}_{-24}$ & \\ [3pt] 
\hline
70110-01-10-00 & 52402.49 & $2.06^{+0.01}_{-0.01}$ & C & This work & $6.6^{+1.7}_{-0.1}$ & $329^{+95}_{-8}$ & \\ [3pt] 
 & & $4.20^{+0.08}_{-0.08}$ & C & \citet{Motta11} & & & see Appendix~\ref{sec:FalseQPOs}, Fig.~\ref{fig:70110-01-10-00}\\ [3pt] 
\hline
70109-04-01-00 & 52405.58 & $5.39^{+0.01}_{-0.01}$ & C & This work & $5.4^{+0.4}_{-0.3}$ & $441^{+46}_{-39}$ & \\ [3pt] 
 & & $5.46^{+0.01}_{-0.01}$ & C & \citet{Motta11} & & & \\ [3pt] 
\hline
70109-04-01-01 & 52405.71 & $5.42^{+0.01}_{-0.00}$ & C & This work & $5.4^{+0.4}_{-0.3}$ & $441^{+46}_{-39}$ & \\ [3pt] 
 & & $5.45^{+0.01}_{-0.01}$ & C & \citet{Motta11} & & & \\ [3pt] 
\hline
70109-04-01-02 & 52406.07 & $5.36^{+0.02}_{-0.02}$ & C & This work & $5.4^{+0.5}_{-0.3}$ & $441^{+51}_{-39}$ & \\ [3pt] 
 & & $5.34^{+0.02}_{-0.02}$ & C & \citet{Motta11} & & & \\ [3pt] 
\hline
70110-01-11-00 & 52406.70 & $5.80^{+0.02}_{-0.02}$ & C & This work & $5.3^{+0.4}_{-0.3}$ & $452^{+42}_{-40}$ & \\ [3pt] 
 & & $5.82^{+0.02}_{-0.02}$ & C & \citet{Motta11} & & & \\ [3pt] 
\hline
70110-01-12-00 & 52410.53 & $7.90^{+0.11}_{-0.11}$ & C & This work & $4.5^{+0.3}_{-0.2}$ & $577^{+47}_{-36}$ & \\ [3pt] 
 & & $8.10^{+0.20}_{-0.20}$ & C & \citet{Motta11} & & & \\ [3pt] 
\hline
70109-01-07-00 & 52411.60 & $5.90^{+0.01}_{-0.01}$ & B & This work & $4.7^{+0.2}_{-0.2}$ & $550^{+26}_{-42}$ & \\ [3pt] 
 & & $5.91^{+0.02}_{-0.02}$ & B & \citet{Gao14} & & & \\ [3pt] 
 & & $5.80^{+0.10}_{-0.10}$ & B & \citet{Motta11} & & & \\ [3pt] 
 & & $7.00^{+0.50}_{-0.50}$ & A & \citet{Motta11} & & & \\ [3pt] 
\hline
70110-01-14-00 & 52416.60 & $6.42^{+0.02}_{-0.02}$ & B & This work & $4.2^{+0.1}_{-0.1}$ & $637^{+30}_{-24}$ & \\ [3pt] 
 & & $6.39^{+0.06}_{-0.06}$ & B & \citet{Gao14} & & & \\ [3pt] 
 & & $6.40^{+0.10}_{-0.10}$ & B & \citet{Motta11} & & & \\ [3pt] 
\hline
70110-01-15-00 & 52419.24 & $5.83^{+0.03}_{-0.03}$ & B & This work & $4.1^{+0.3}_{-0.1}$ & $668^{+62}_{-25}$ & \\ [3pt] 
 & & $5.82^{+0.03}_{-0.03}$ & B & \citet{Gao14} & & & \\ [3pt] 
 & & $5.70^{+0.03}_{-0.03}$ & B & \citet{Motta11} & & & \\ [3pt] 
\hline
70108-03-02-00 & 52419.40 & $5.20^{+0.02}_{-0.02}$ & B & \citet{Gao14} & $3.5^{+0.1}_{-0.1}$ & $854^{+41}_{-21}$ & \\ [3pt] 
 & & $5.60^{+0.10}_{-0.10}$ & B & \citet{Motta11} & & & \\ [3pt] 
 & & $6.70^{+0.50}_{-0.50}$ & A & \citet{Motta11} & & & \\ [3pt] 
 & & $6.80^{+0.40}_{-0.40}$ & A & \citet{Motta11} & & & \\ [3pt] 
\hline
70109-01-20-00 & 52504.71 & $4.76^{+0.09}_{-0.08}$ & B & This work & $2.4^{+0.1}_{-0.1}$ & $1498^{+54}_{-134}$ & \\ [3pt] 
\hline
70110-01-45-00 & 52524.95 & $7.04^{+0.27}_{-0.30}$ & A & This work & $4.0^{+0.2}_{-0.1}$ & $702^{+58}_{-35}$ & \\ [3pt] 
 & & $7.20^{+0.60}_{-0.60}$ & A & \citet{Motta11} & & & \\ [3pt] 
\hline
70109-01-23-00 & 52529.58 & $7.40^{+1.30}_{-1.30}$ & A & \citet{Motta11} & $3.7^{+0.2}_{-0.1}$ & $774^{+46}_{-39}$ & \\ [3pt] 
\hline
70110-01-47-00 & 52532.75 & $5.71^{+0.02}_{-0.02}$ & B & This work & $3.8^{+0.2}_{-0.1}$ & $737^{+60}_{-37}$ & \\ [3pt] 
 & & $5.75^{+0.03}_{-0.04}$ & B & \citet{Gao14} & & & \\ [3pt] 
 & & $6.20^{+0.10}_{-0.10}$ & B & \citet{Motta11} & & & \\ [3pt] 
\hline
70109-01-24-00 & 52536.36 & $8.00^{+0.90}_{-0.90}$ & A & \citet{Motta11} & $3.6^{+0.1}_{-0.1}$ & $813^{+29}_{-41}$ & \\ [3pt] 
\hline
70109-01-37-00 & 52694.92 & $8.60^{+0.20}_{-0.20}$ & C & \citet{Motta11} & $2.8^{+0.1}_{-0.1}$ & $1202^{+85}_{-76}$ & \\ [3pt] 
\hline
70128-02-02-00 & 52696.36 & $8.02^{+0.03}_{-0.03}$ & C & This work & $2.9^{+0.1}_{-0.1}$ & $1117^{+79}_{-42}$ & \\ [3pt] 
 & & $8.02^{+0.04}_{-0.04}$ & C & \citet{Motta11} & & & \\ [3pt] 
\hline
50117-01-03-01 & 52706.77 & $6.70^{+0.10}_{-0.10}$ & C & \citet{Motta11} & $3.3^{+0.4}_{-0.2}$ & $919^{+145}_{-82}$ & \\ [3pt] 
\hline
50117-01-03-00 & 52706.84 & $6.77^{+0.02}_{-0.02}$ & C & \citet{Motta11} & $3.3^{+0.6}_{-0.1}$ & $941^{+204}_{-59}$ & \\ [3pt] 
\hline
70110-01-89-00 & 52707.91 & $0.88^{+0.02}_{-0.02}$ & B & This work & $3.0^{+0.1}_{-0.1}$ & $1090^{+52}_{-83}$ & \\ [3pt] 
 & & $0.86^{+0.04}_{-0.04}$ & B & \citet{Gao14} & & & \\ [3pt] 
 & & $0.90^{+0.05}_{-0.05}$ & B & \citet{Motta11} & & & \\ [3pt] 
\hline
70109-02-01-00 & 52709.86 & $8.05^{+0.07}_{-0.07}$ & C & This work & $3.0^{+0.2}_{-0.1}$ & $1090^{+89}_{-55}$ & \\ [3pt] 
 & & $8.00^{+0.10}_{-0.10}$ & C & \citet{Motta11} & & & \\ [3pt] 
\hline
70109-02-01-01 & 52709.99 & $8.10^{+0.10}_{-0.10}$ & C & \citet{Motta11} & $2.9^{+0.1}_{-0.1}$ & $1117^{+53}_{-42}$ & \\ [3pt] 
\hline
60705-01-56-00 & 52710.71 & $7.80^{+0.10}_{-0.10}$ & C & \citet{Motta11} & $3.1^{+0.2}_{-0.1}$ & $1038^{+85}_{-39}$ & \\ [3pt] 
\hline
70110-01-94-00 & 52724.23 & $6.19^{+0.03}_{-0.03}$ & C & This work & $3.4^{+0.7}_{-0.2}$ & $875^{+206}_{-67}$ & \\ [3pt] 
 & & $6.10^{+0.10}_{-0.10}$ & C & \citet{Motta11} & & & \\ [3pt] 
\hline
70110-01-95-00 & 52727.25 & $4.70^{+0.10}_{-0.10}$ & C & \citet{Motta11} & $4.0^{+2.8}_{-0.5}$ & $685^{+372}_{-148}$ & \\ [3pt] 
\hline
60705-01-59-00 & 52731.56 & $2.90^{+0.10}_{-0.10}$ & C & \citet{Motta11} & $6.3^{+23.0}_{-2.0}$ & $354^{+319}_{-275}$ & \\ [3pt] 
\hline
60705-01-68-00 & 53218.11 & $0.50^{+0.00}_{-0.00}$ & C & \citet{Motta11} & $27.6^{+301.8}_{-19.3}$ & $38^{+37}_{-193}$ & \\ [3pt] 
\hline
60705-01-68-01 & 53222.25 & $1.04^{+0.02}_{-0.01}$ & C & This work & $15.6^{+210.8}_{-8.8}$ & $90^{+88}_{-223}$ & \\ [3pt] 
 & & $1.03^{+0.04}_{-0.04}$ & C & \citet{Motta11} & & & \\ [3pt] 
\hline
60705-01-69-00 & 53225.40 & $1.27^{+0.02}_{-0.02}$ & C & This work & $12.2^{+191.4}_{-5.6}$ & $130^{+128}_{-199}$ & \\ [3pt] 
 & & $1.30^{+0.00}_{-0.00}$ & C & \citet{Motta11} & & & \\ [3pt] 
\hline
90704-01-01-00 & 53226.43 & $2.00^{+0.10}_{-0.10}$ & C & \citet{Motta11} & $12.2^{+171.0}_{-5.8}$ & $130^{+128}_{-211}$ & \\ [3pt] 
\hline
60705-01-69-01 & 53228.99 & $2.90^{+0.20}_{-0.20}$ & C & \citet{Motta11} & $7.4^{+23.8}_{-1.7}$ & $277^{+245}_{-138}$ & \\ [3pt] 
\hline
60705-01-70-00 & 53230.96 & $4.33^{+0.02}_{-0.03}$ & C & This work & $5.7^{+0.0}_{-0.0}$ & $410^{+0}_{-0}$ & \\ [3pt] 
 & & $4.30^{+0.10}_{-0.10}$ & C & \citet{Motta11} & & & \\ [3pt] 
\hline
90110-02-01-01 & 53232.01 & $5.05^{+0.04}_{-0.06}$ & C & This work & $4.2^{+0.6}_{-0.1}$ & $652^{+122}_{-33}$ & \\ [3pt] 
\hline
90110-02-01-02 & 53232.34 & $5.25^{+0.05}_{-0.04}$ & C & This work & $5.1^{+0.5}_{-0.6}$ & $475^{+65}_{-89}$ & \\ [3pt] 
 & & $5.20^{+0.10}_{-0.10}$ & C & \citet{Motta11} & & & \\ [3pt] 
\hline
90110-02-01-00 & 53232.40 & $5.80^{+0.10}_{-0.10}$ & C & \citet{Motta11} & $4.9^{+0.3}_{-0.5}$ & $511^{+48}_{-81}$ & \\ [3pt] 
\hline
90110-02-01-03 & 53232.99 & $4.10^{+0.01}_{-0.01}$ & B & This work & $3.4^{+0.1}_{-0.1}$ & $875^{+32}_{-22}$ & \\ [3pt] 
 & & $4.08^{+0.02}_{-0.02}$ & B & \citet{Gao14} & & & \\ [3pt] 
 & & $4.10^{+0.04}_{-0.04}$ & B & \citet{Motta11} & & & \\ [3pt] 
\hline
90704-01-02-00 & 53233.39 & $4.21^{+0.01}_{-0.01}$ & B & This work & $3.5^{+0.1}_{-0.1}$ & $833^{+30}_{-21}$ & \\ [3pt] 
 & & $4.21^{+0.02}_{-0.02}$ & B & \citet{Gao14} & & & \\ [3pt] 
 & & $4.40^{+0.20}_{-0.20}$ & B & \citet{Motta11} & & & \\ [3pt] 
\hline
60705-01-84-02 & 53333.90 & $4.97^{+0.02}_{-0.02}$ & B & This work & $3.8^{+0.2}_{-0.1}$ & $755^{+53}_{-28}$ & \\ [3pt] 
 & & $4.96^{+0.03}_{-0.03}$ & B & \citet{Gao14} & & & \\ [3pt] 
 & & $5.20^{+0.10}_{-0.10}$ & B & \citet{Motta11} & & & \\ [3pt] 
\hline
91105-04-10-00 & 53466.75 & $3.41^{+0.05}_{-0.04}$ & B & This work & $3.1^{+0.3}_{-0.1}$ & $1013^{+117}_{-64}$ & \\ [3pt] 
 & & $3.38^{+0.11}_{-0.10}$ & B & \citet{Gao14} & & & \\ [3pt] 
 & & $3.40^{+0.10}_{-0.10}$ & B & \citet{Motta11} & & & \\ [3pt] 
\hline
90704-01-11-00 & 53472.33 & $2.70^{+0.10}_{-0.10}$ & C & \citet{Motta11} & $5.2^{+18.2}_{-0.9}$ & $463^{+414}_{-143}$ & \\ [3pt] 
\hline
92035-01-01-01 & 54128.94 & $0.14^{+0.01}_{-0.01}$ & C & \citet{2017ApJ...845..143Z} & $13.7^{+246.4}_{-5.7}$ & $109^{+108}_{-136}$ & \\ [3pt] 
\hline
92035-01-01-03 & 54130.13 & $0.17^{+0.01}_{-0.01}$ & C & \citet{2017ApJ...845..143Z} & $18.4^{+12.6}_{-7.8}$ & $70^{+38}_{-91}$ & \\ [3pt] 
\hline
92035-01-01-02 & 54131.11 & $0.18^{+0.01}_{-0.01}$ & C & \citet{2017ApJ...845..143Z} & $18.7^{+12.3}_{-7.9}$ & $69^{+37}_{-89}$ & \\ [3pt] 
\hline
92035-01-01-04 & 54132.09 & $0.20^{+0.01}_{-0.01}$ & C & \citet{2017ApJ...845..143Z} & $20.9^{+11.6}_{-9.7}$ & $58^{+28}_{-89}$ & \\ [3pt] 
\hline
92035-01-02-00 & 54133.00 & $0.23^{+0.01}_{-0.01}$ & C & \citet{2017ApJ...845..143Z} & $19.6^{+11.8}_{-8.3}$ & $64^{+32}_{-83}$ & \\ [3pt] 
\hline
92035-01-02-01 & 54133.92 & $0.26^{+0.00}_{-0.00}$ & C & This work & $19.0^{+12.0}_{-6.8}$ & $67^{+35}_{-63}$ & \\ [3pt] 
 & & $0.28^{+0.01}_{-0.01}$ & C & \citet{Motta11} & & & \\ [3pt] 
 & & $0.26^{+0.01}_{-0.01}$ & C & \citet{2017ApJ...845..143Z} & & & \\ [3pt] 
\hline
92035-01-02-02 & 54135.03 & $0.30^{+0.01}_{-0.01}$ & C & \citet{Motta11} & $19.9^{+12.0}_{-7.1}$ & $62^{+32}_{-58}$ & \\ [3pt] 
 & & $0.29^{+0.01}_{-0.01}$ & C & \citet{2017ApJ...845..143Z} & & & \\ [3pt] 
\hline
92035-01-02-03 & 54136.01 & $0.37^{+0.01}_{-0.01}$ & C & \citet{Motta11} & $28.0^{+9.0}_{-12.3}$ & $37^{+13}_{-52}$ & \\ [3pt] 
 & & $0.36^{+0.01}_{-0.01}$ & C & \citet{2017ApJ...845..143Z} & & & \\ [3pt] 
\hline
92035-01-02-04 & 54137.00 & $0.43^{+0.01}_{-0.01}$ & C & \citet{Motta11} & $9.9^{+163.1}_{-1.5}$ & $179^{+176}_{-49}$ & \\ [3pt] 
 & & $0.43^{+0.01}_{-0.01}$ & C & \citet{2017ApJ...845..143Z} & & & \\ [3pt] 
\hline
92035-01-02-08 & 54137.85 & $0.55^{+0.02}_{-0.02}$ & C & \citet{Motta11} & $9.7^{+143.4}_{-1.3}$ & $183^{+180}_{-45}$ & \\ [3pt] 
 & & $0.54^{+0.01}_{-0.01}$ & C & \citet{2017ApJ...845..143Z} & & & \\ [3pt] 
\hline
92035-01-02-07 & 54138.83 & $0.90^{+0.01}_{-0.01}$ & C & \citet{Motta11} & $9.0^{+121.1}_{-0.8}$ & $207^{+203}_{-33}$ & \\ [3pt] 
 & & $0.89^{+0.01}_{-0.01}$ & C & \citet{2017ApJ...845..143Z} & & & \\ [3pt] 
\hline
92035-01-02-06 & 54139.94 & $0.99^{+0.01}_{-0.01}$ & C & \citet{Motta11} & $7.9^{+14.8}_{-0.4}$ & $251^{+200}_{-19}$ & \\ [3pt] 
 & & $0.99^{+0.01}_{-0.01}$ & C & \citet{2017ApJ...845..143Z} & & & \\ [3pt] 
\hline
92035-01-03-00 & 54140.20 & $1.13^{+0.01}_{-0.01}$ & C & \citet{Motta11} & $7.6^{+10.2}_{-0.3}$ & $264^{+190}_{-17}$ & \\ [3pt] 
 & & $1.13^{+0.01}_{-0.01}$ & C & \citet{2017ApJ...845..143Z} & & & \\ [3pt] 
\hline
92035-01-03-01 & 54141.06 & $1.68^{+0.01}_{-0.01}$ & C & \citet{Motta11} & $7.1^{+3.9}_{-0.2}$ & $291^{+141}_{-15}$ & \\ [3pt] 
 & & $1.68^{+0.01}_{-0.01}$ & C & \citet{2017ApJ...845..143Z} & & & \\ [3pt] 
\hline
92035-01-03-02 & 54142.04 & $2.45^{+0.01}_{-0.01}$ & C & \citet{Motta11} & $6.6^{+0.4}_{-0.1}$ & $329^{+31}_{-8}$ & \\ [3pt] 
 & & $2.45^{+0.01}_{-0.01}$ & C & \citet{2017ApJ...845..143Z} & & & \\ [3pt] 
\hline
92035-01-03-03 & 54143.02 & $3.52^{+0.01}_{-0.01}$ & C & \citet{Motta11} & $6.3^{+0.5}_{-0.4}$ & $354^{+37}_{-41}$ & \\ [3pt] 
 & & $3.52^{+0.01}_{-0.01}$ & C & \citet{2017ApJ...845..143Z} & & & \\ [3pt] 
\hline
92428-01-04-00 & 54143.87 & $4.34^{+0.01}_{-0.01}$ & C & This work & $6.0^{+0.5}_{-0.4}$ & $381^{+44}_{-44}$ & \\ [3pt] 
 & & $4.34^{+0.02}_{-0.02}$ & C & \citet{Motta11} & & & \\ [3pt] 
 & & $4.34^{+0.01}_{-0.01}$ & C & \citet{2017ApJ...845..143Z} & & & \\ [3pt] 
\hline
92428-01-04-01 & 54143.95 & $4.24^{+0.01}_{-0.01}$ & C & This work & $6.0^{+0.6}_{-0.3}$ & $381^{+48}_{-34}$ & \\ [3pt] 
 & & $4.23^{+0.02}_{-0.02}$ & C & \citet{Motta11} & & & \\ [3pt] 
 & & $4.24^{+0.01}_{-0.01}$ & C & \citet{2017ApJ...845..143Z} & & & \\ [3pt] 
\hline
92428-01-04-02 & 54144.09 & $4.14^{+0.01}_{-0.01}$ & C & This work & $6.1^{+0.5}_{-0.4}$ & $372^{+43}_{-43}$ & \\ [3pt] 
 & & $4.13^{+0.03}_{-0.03}$ & C & \citet{Motta11} & & & \\ [3pt] 
 & & $4.14^{+0.01}_{-0.01}$ & C & \citet{2017ApJ...845..143Z} & & & \\ [3pt] 
\hline
92428-01-04-03 & 54144.87 & $4.99^{+0.01}_{-0.01}$ & C & This work & $5.8^{+0.5}_{-0.4}$ & $400^{+46}_{-47}$ & \\ [3pt] 
 & & $4.99^{+0.03}_{-0.03}$ & C & \citet{Motta11} & & & \\ [3pt] 
 & & $4.98^{+0.01}_{-0.01}$ & C & \citet{2017ApJ...845..143Z} & & & \\ [3pt] 
\hline
92035-01-03-05 & 54145.11 & $5.80^{+0.03}_{-0.03}$ & C & \citet{Motta11} & $5.4^{+0.5}_{-0.3}$ & $441^{+51}_{-39}$ & \\ [3pt] 
 & & $5.80^{+0.02}_{-0.02}$ & C & \citet{2017ApJ...845..143Z} & & & \\ [3pt] 
\hline
92428-01-04-04 & 54145.96 & $5.58^{+0.02}_{-0.02}$ & C & This work & $5.6^{+0.5}_{-0.4}$ & $420^{+48}_{-49}$ & \\ [3pt] 
 & & $5.61^{+0.02}_{-0.02}$ & C & \citet{2017ApJ...845..143Z} & & & \\ [3pt] 
\hline
92035-01-03-06 & 54146.03 & $5.55^{+0.01}_{-0.01}$ & C & \citet{2017ApJ...845..143Z} & $5.6^{+0.5}_{-0.4}$ & $420^{+48}_{-43}$ & \\ [3pt] 
\hline
92035-01-04-00 & 54147.01 & $6.70^{+0.04}_{-0.04}$ & B & This work & $4.6^{+0.3}_{-0.1}$ & $563^{+46}_{-28}$ & \\ [3pt] 
 & & $6.70^{+0.02}_{-0.02}$ & B & \citet{Gao14} & & & \\ [3pt] 
 & & $6.70^{+0.20}_{-0.20}$ & B & \citet{Motta11} & & & \\ [3pt] 
\hline
92085-01-02-06 & 54160.90 & $7.80^{+0.70}_{-0.70}$ & A & \citet{Motta11} & $3.7^{+0.1}_{-0.1}$ & $774^{+37}_{-29}$ & \\ [3pt] 
\hline
92085-01-03-00 & 54161.67 & $7.72^{+0.18}_{-0.18}$ & B & This work & $4.7^{+0.4}_{-0.1}$ & $550^{+57}_{-21}$ & \\ [3pt] 
 & & $3.96^{+0.39}_{-0.39}$ & B & This work & & & see Appendix~\ref{sec:FalseQPOs}, Fig.~\ref{fig:92085-01-03-00} \\ [3pt] 
 & & $7.10^{+0.10}_{-0.10}$ & C & \citet{Motta11} & & & \\ [3pt] 
\hline
92085-01-03-01 & 54162.66 & $6.27^{+0.06}_{-0.05}$ & B & This work & $4.3^{+0.3}_{-0.1}$ & $621^{+58}_{-23}$ & \\ [3pt] 
 & & $6.22^{+0.02}_{-0.02}$ & B & \citet{Gao14} & & & \\ [3pt] 
 & & $6.40^{+0.10}_{-0.10}$ & B & \citet{Motta11} & & & \\ [3pt] 
\hline
92085-01-03-02 & 54163.70 & $7.30^{+0.20}_{-0.20}$ & C & \citet{Motta11} & $4.5^{+0.3}_{-0.1}$ & $577^{+60}_{-22}$ & \\ [3pt] 
\hline
92085-01-03-03 & 54164.56 & $7.39^{+0.21}_{-0.22}$ & B & This work & $4.6^{+0.3}_{-0.2}$ & $563^{+52}_{-36}$ & \\ [3pt] 
 & & $7.00^{+0.20}_{-0.20}$ & C & \citet{Motta11} & & & see Appendix~\ref{sec:FalseQPOs}, Fig.~\ref{fig:92085-01-03-03} \\ [3pt] 
\hline
92085-01-03-04 & 54165.53 & $7.70^{+0.70}_{-0.70}$ & A & \citet{Motta11} & $3.6^{+0.1}_{-0.1}$ & $813^{+39}_{-30}$ & \\ [3pt] 
\hline
92704-03-10-00 & 54231.60 & $1.00^{+0.06}_{-0.06}$ & B & \citet{Gao14} & $2.9^{+0.1}_{-0.1}$ & $1117^{+53}_{-42}$ & \\ [3pt] 
 & & $1.00^{+0.10}_{-0.10}$ & B & \citet{Motta11} & & & \\ [3pt] 
\hline
92704-03-10-11 & 54232.60 & $1.69^{+0.07}_{-0.07}$ & B & \citet{Gao14} & $3.0^{+0.1}_{-0.0}$ & $1090^{+39}_{-27}$ & \\ [3pt] 
 & & $1.70^{+0.10}_{-0.10}$ & B & \citet{Motta11} & & & \\ [3pt] 
\hline
92704-03-10-12 & 54233.60 & $1.73^{+0.06}_{-0.06}$ & B & \citet{Gao14} & $3.2^{+0.2}_{-0.1}$ & $989^{+70}_{-24}$ & \\ [3pt] 
 & & $1.80^{+0.20}_{-0.20}$ & B & \citet{Motta11} & & & \\ [3pt] 
\hline
92704-03-11-00 & 54234.84 & $4.00^{+0.10}_{-0.10}$ & C & \citet{Motta11} & $4.2^{+3.8}_{-0.4}$ & $652^{+407}_{-103}$ & \\ [3pt] 
\hline
92704-03-11-01 & 54235.79 & $3.30^{+0.10}_{-0.10}$ & C & \citet{Motta11} & $5.8^{+7.4}_{-1.6}$ & $400^{+284}_{-260}$ & \\ [3pt] 
\hline
92704-04-01-01 & 54236.45 & $3.00^{+0.20}_{-0.20}$ & C & \citet{Motta11} & $5.2^{+15.0}_{-1.0}$ & $463^{+402}_{-166}$ & \\ [3pt] 
\hline
92704-04-01-02 & 54236.51 & $2.70^{+0.10}_{-0.10}$ & C & \citet{Motta11} & $6.3^{+19.4}_{-1.7}$ & $354^{+311}_{-209}$ & \\ [3pt] 
\hline
92704-03-12-00 & 54236.59 & $2.70^{+0.40}_{-0.40}$ & C & \citet{Motta11} & $6.3^{+18.2}_{-1.8}$ & $354^{+308}_{-230}$ & \\ [3pt] 
\hline
92704-04-01-04 & 54237.36 & $2.80^{+0.10}_{-0.10}$ & C & \citet{Motta11} & $6.0^{+0.6}_{-0.7}$ & $381^{+52}_{-82}$ & \\ [3pt] 
\hline
92704-04-01-05 & 54237.42 & $3.10^{+0.30}_{-0.30}$ & C & \citet{Motta11} & $5.8^{+0.8}_{-0.5}$ & $400^{+71}_{-58}$ & \\ [3pt] 
\hline
92704-03-12-01 & 54237.49 & $2.70^{+0.10}_{-0.10}$ & C & \citet{Motta11} & $5.9^{+0.8}_{-0.6}$ & $390^{+65}_{-73}$ & \\ [3pt] 
\hline
95409-01-04-00 & 55225.71 & $2.48^{+0.02}_{-0.02}$ & C & This work & $29.0^{+28.0}_{-5.9}$ & $36^{+23}_{-15}$ & \\ [3pt] 
\hline
95409-01-11-02 & 55277.48 & $0.10^{+0.01}_{-0.01}$ & C & \citet{Nandi12} & $30.9^{+24.2}_{-14.1}$ & $32^{+19}_{-48}$ & \\ [3pt] 
\hline
95409-01-11-03 & 55279.57 & $0.11^{+0.01}_{-0.01}$ & C & \citet{Nandi12} & $29.4^{+23.9}_{-12.5}$ & $35^{+20}_{-45}$ & \\ [3pt] 
\hline
95409-01-12-00 & 55281.59 & $0.13^{+0.01}_{-0.01}$ & C & \citet{Nandi12} & $34.1^{+21.1}_{-16.3}$ & $28^{+14}_{-46}$ & \\ [3pt] 
\hline
95409-01-12-04 & 55286.73 & $0.22^{+0.01}_{-0.01}$ & C & \citet{Motta11} & $36.4^{+18.3}_{-13.3}$ & $25^{+12}_{-25}$ & \\ [3pt] 
\hline
95409-01-12-03 & 55287.60 & $0.20^{+0.00}_{-0.00}$ & C & \citet{Nandi12} & $33.6^{+21.6}_{-13.1}$ & $29^{+15}_{-32}$ & \\ [3pt] 
\hline
95409-01-13-03 & 55288.37 & $0.20^{+0.10}_{-0.10}$ & C & \citet{Motta11} & $37.6^{+14.9}_{-17.7}$ & $24^{+9}_{-38}$ & \\ [3pt] 
\hline
95409-01-13-00 & 55289.62 & $0.27^{+0.00}_{-0.00}$ & C & This work & $29.4^{+16.7}_{-11.8}$ & $35^{+17}_{-40}$ & \\ [3pt] 
 & & $0.26^{+0.01}_{-0.01}$ & C & \citet{Motta11} & & & \\ [3pt] 
 & & $0.26^{+0.01}_{-0.00}$ & C & \citet{Nandi12} & & & \\ [3pt] 
\hline
95409-01-13-04 & 55290.72 & $0.29^{+0.01}_{-0.01}$ & C & \citet{Motta11} & $35.2^{+10.5}_{-17.3}$ & $27^{+9}_{-47}$ & \\ [3pt] 
\hline
95409-01-13-02 & 55291.65 & $0.32^{+0.00}_{-0.00}$ & C & This work & $33.6^{+14.1}_{-13.6}$ & $29^{+12}_{-34}$ & \\ [3pt] 
 & & $0.32^{+0.01}_{-0.01}$ & C & \citet{Motta11} & & & \\ [3pt] 
 & & $0.32^{+0.00}_{-0.01}$ & C & \citet{Nandi12} & & & \\ [3pt] 
\hline
95409-01-13-05 & 55292.78 & $0.37^{+0.01}_{-0.01}$ & C & This work & $34.7^{+8.9}_{-17.3}$ & $27^{+8}_{-50}$ & \\ [3pt] 
 & & $0.38^{+0.02}_{-0.02}$ & C & \citet{Motta11} & & & \\ [3pt] 
 & & $0.36^{+0.00}_{-0.00}$ & C & \citet{Nandi12} & & & \\ [3pt] 
\hline
95409-01-13-01 & 55293.09 & $0.37^{+0.02}_{-0.02}$ & C & This work & $28.5^{+11.0}_{-11.3}$ & $36^{+14}_{-41}$ & \\ [3pt] 
 & & $0.38^{+0.05}_{-0.05}$ & C & \citet{Motta11} & & & \\ [3pt] 
\hline
95409-01-13-06 & 55294.12 & $0.47^{+0.02}_{-0.02}$ & C & \citet{Motta11} & $29.0^{+8.9}_{-13.1}$ & $36^{+12}_{-52}$ & \\ [3pt] 
\hline
95409-01-14-00 & 55295.00 & $0.55^{+0.00}_{-0.00}$ & C & This work & $9.6^{+32.9}_{-1.9}$ & $187^{+167}_{-73}$ & \\ [3pt] 
 & & $0.55^{+0.01}_{-0.01}$ & C & \citet{Nandi12} & & & \\ [3pt] 
\hline
95409-01-14-01 & 55296.25 & $1.05^{+0.01}_{-0.01}$ & C & This work & $8.4^{+13.0}_{-1.0}$ & $228^{+172}_{-46}$ & \\ [3pt] 
 & & $1.04^{+0.01}_{-0.01}$ & C & \citet{Motta11} & & & \\ [3pt] 
 & & $1.03^{+0.01}_{-0.01}$ & C & \citet{Nandi12} & & & \\ [3pt] 
\hline
95409-01-14-02 & 55297.87 & $1.25^{+0.01}_{-0.01}$ & C & \citet{Motta11} & $7.2^{+12.5}_{-0.3}$ & $284^{+221}_{-22}$ & \\ [3pt] 
 & & $1.24^{+0.01}_{-0.01}$ & C & \citet{Nandi12} & & & \\ [3pt] 
\hline
95409-01-14-03 & 55298.70 & $1.59^{+0.01}_{-0.01}$ & C & \citet{Motta11} & $6.9^{+7.0}_{-0.2}$ & $306^{+199}_{-15}$ & \\ [3pt] 
 & & $1.59^{+0.01}_{-0.01}$ & C & \citet{Nandi12} & & & \\ [3pt] 
\hline
95409-01-14-06 & 55299.77 & $2.43^{+0.01}_{-0.01}$ & C & \citet{Motta11} & $6.3^{+2.6}_{-0.2}$ & $354^{+142}_{-22}$ & \\ [3pt] 
 & & $2.42^{+0.01}_{-0.01}$ & C & \citet{Nandi12} & & & \\ [3pt] 
\hline
95409-01-14-04 & 55300.34 & $2.38^{+0.01}_{-0.01}$ & C & This work & $6.3^{+2.1}_{-0.3}$ & $354^{+126}_{-32}$ & \\ [3pt] 
 & & $2.38^{+0.01}_{-0.01}$ & C & \citet{Motta11} & & & \\ [3pt] 
 & & $2.37^{+0.01}_{-0.01}$ & C & \citet{Nandi12} & & & \\ [3pt] 
\hline
95409-01-14-07 & 55300.92 & $2.92^{+0.00}_{-0.00}$ & C & \citet{Motta11} & $6.1^{+1.2}_{-0.3}$ & $372^{+91}_{-33}$ & \\ [3pt] 
\hline
95409-01-14-05 & 55301.79 & $3.65^{+0.01}_{-0.01}$ & C & This work & $5.8^{+0.6}_{-0.4}$ & $400^{+55}_{-47}$ & \\ [3pt] 
 & & $3.64^{+0.02}_{-0.02}$ & C & \citet{Motta11} & & & \\ [3pt] 
 & & $3.64^{+0.01}_{-0.01}$ & C & \citet{Nandi12} & & & \\ [3pt] 
\hline
95409-01-15-00 & 55302.20 & $4.15^{+0.01}_{-0.01}$ & C & This work & $5.6^{+0.3}_{-0.4}$ & $420^{+34}_{-55}$ & \\ [3pt] 
 & & $4.15^{+0.03}_{-0.03}$ & C & \citet{Motta11} & & & \\ [3pt] 
 & & $4.18^{+0.04}_{-0.07}$ & C & \citet{Nandi12} & & & \\ [3pt] 
\hline
95409-01-15-01 & 55303.61 & $5.65^{+0.02}_{-0.02}$ & C & This work & $5.0^{+0.4}_{-0.2}$ & $499^{+52}_{-31}$ & \\ [3pt] 
 & & $5.65^{+0.04}_{-0.04}$ & C & \citet{Motta11} & & & \\ [3pt] 
 & & $5.69^{+0.03}_{-0.03}$ & C & \citet{Nandi12} & & & \\ [3pt] 
\hline
95409-01-15-02 & 55304.71 & $5.72^{+0.02}_{-0.02}$ & B & This work & $4.2^{+0.3}_{-0.1}$ & $652^{+61}_{-24}$ & \\ [3pt] 
 & & $5.73^{+0.03}_{-0.03}$ & B & \citet{Gao14} & & & \\ [3pt] 
 & & $5.60^{+0.10}_{-0.10}$ & B & \citet{Motta11} & & & \\ [3pt] 
 & & $5.74^{+0.03}_{-0.03}$ & B & \citet{Nandi12} & & & \\ [3pt] 
\hline
95409-01-15-06 & 55308.98 & $5.68^{+0.01}_{-0.01}$ & B & This work & $3.9^{+0.2}_{-0.1}$ & $719^{+43}_{-27}$ & \\ [3pt] 
 & & $5.68^{+0.02}_{-0.02}$ & B & \citet{Gao14} & & & \\ [3pt] 
 & & $5.90^{+0.20}_{-0.20}$ & B & \citet{Motta11} & & & \\ [3pt] 
 & & $5.68^{+0.02}_{-0.02}$ & B & \citet{Nandi12} & & & \\ [3pt] 
\hline
95409-01-16-05 & 55315.70 & $5.94^{+0.02}_{-0.02}$ & B & This work & $4.2^{+0.2}_{-0.1}$ & $652^{+39}_{-33}$ & \\ [3pt] 
 & & $5.93^{+0.03}_{-0.03}$ & B & \citet{Gao14} & & & \\ [3pt] 
 & & $6.10^{+0.20}_{-0.20}$ & B & \citet{Motta11} & & & \\ [3pt] 
 & & $5.91^{+0.04}_{-0.04}$ & B & \citet{Nandi12} & & & \\ [3pt] 
\hline
95409-01-17-00 & 55316.11 & $5.97^{+0.03}_{-0.03}$ & B & This work & $4.1^{+0.2}_{-0.1}$ & $668^{+40}_{-25}$ & \\ [3pt] 
 & & $5.99^{+0.05}_{-0.05}$ & B & \citet{Gao14} & & & \\ [3pt] 
 & & $5.90^{+0.10}_{-0.10}$ & B & \citet{Motta11} & & & \\ [3pt] 
 & & $6.07^{+0.08}_{-0.08}$ & B & \citet{Nandi12} & & & \\ [3pt] 
\hline
95409-01-17-02 & 55318.44 & $6.66^{+0.12}_{-0.13}$ & C & This work & $4.4^{+0.4}_{-0.1}$ & $606^{+70}_{-23}$ & \\ [3pt] 
 & & $6.67^{+0.19}_{-0.19}$ & C & \citet{Motta11} & & & \\ [3pt] 
 & & $6.65^{+0.12}_{-0.11}$ & B & \citet{Nandi12} & & & see Appendix~\ref{sec:FalseQPOs}, Fig.~\ref{fig:95409-01-17-02} \\ [3pt] 
\hline
95409-01-17-05 & 55321.72 & $5.24^{+0.01}_{-0.01}$ & B & This work & $3.8^{+0.2}_{-0.1}$ & $755^{+53}_{-28}$ & \\ [3pt] 
 & & $5.23^{+0.02}_{-0.02}$ & B & \citet{Gao14} & & & \\ [3pt] 
 & & $5.30^{+0.10}_{-0.10}$ & B & \citet{Motta11} & & & \\ [3pt] 
 & & $5.25^{+0.02}_{-0.02}$ & B & \citet{Nandi12} & & & \\ [3pt] 
\hline
95409-01-17-06 & 55322.23 & $5.20^{+0.02}_{-0.02}$ & B & This work & $3.7^{+0.2}_{-0.1}$ & $774^{+46}_{-29}$ & \\ [3pt] 
 & & $5.11^{+0.02}_{-0.03}$ & B & \citet{Gao14} & & & \\ [3pt] 
 & & $5.20^{+0.10}_{-0.10}$ & B & \citet{Motta11} & & & \\ [3pt] 
 & & $5.11^{+0.03}_{-0.03}$ & B & \citet{Nandi12} & & & \\ [3pt] 
\hline
95409-01-18-00 & 55323.21 & $5.39^{+0.02}_{-0.01}$ & B & This work & $4.0^{+0.3}_{-0.2}$ & $685^{+71}_{-52}$ & \\ [3pt] 
 & & $5.37^{+0.02}_{-0.02}$ & B & \citet{Gao14} & & & \\ [3pt] 
 & & $5.50^{+0.10}_{-0.10}$ & B & \citet{Motta11} & & & \\ [3pt] 
 & & $5.39^{+0.02}_{-0.03}$ & B & \citet{Nandi12} & & & \\ [3pt] 
\hline
95335-01-01-07 & 55324.19 & $5.33^{+0.02}_{-0.02}$ & B & This work & $3.8^{+0.2}_{-0.1}$ & $737^{+44}_{-18}$ & \\ [3pt] 
 & & $5.33^{+0.03}_{-0.03}$ & B & \citet{Gao14} & & & \\ [3pt] 
 & & $5.30^{+0.04}_{-0.04}$ & B & \citet{Motta11} & & & \\ [3pt] 
\hline
95335-01-01-00 & 55324.25 & $5.26^{+0.01}_{-0.01}$ & B & This work & $3.8^{+0.3}_{-0.1}$ & $737^{+69}_{-28}$ & \\ [3pt] 
 & & $5.22^{+0.02}_{-0.02}$ & B & \citet{Gao14} & & & \\ [3pt] 
 & & $5.30^{+0.03}_{-0.03}$ & B & \citet{Motta11} & & & \\ [3pt] 
\hline
95335-01-01-01 & 55324.39 & $5.09^{+0.01}_{-0.01}$ & B & This work & $3.8^{+0.2}_{-0.2}$ & $755^{+45}_{-48}$ & \\ [3pt] 
 & & $5.09^{+0.02}_{-0.02}$ & B & \citet{Gao14} & & & \\ [3pt] 
 & & $5.10^{+0.02}_{-0.02}$ & B & \citet{Motta11} & & & \\ [3pt] 
\hline
95335-01-01-05 & 55326.18 & $4.99^{+0.02}_{-0.02}$ & B & This work & $3.6^{+0.1}_{-0.1}$ & $813^{+39}_{-30}$ & \\ [3pt] 
 & & $4.89^{+0.02}_{-0.02}$ & B & \citet{Gao14} & & & \\ [3pt] 
 & & $4.90^{+0.03}_{-0.03}$ & B & \citet{Motta11} & & & \\ [3pt] 
\hline
95335-01-01-06 & 55326.28 & $4.89^{+0.02}_{-0.02}$ & B & This work & $3.5^{+0.1}_{-0.1}$ & $854^{+41}_{-32}$ & \\ [3pt] 
 & & $4.91^{+0.03}_{-0.03}$ & B & \citet{Gao14} & & & \\ [3pt] 
 & & $4.90^{+0.03}_{-0.03}$ & B & \citet{Motta11} & & & \\ [3pt] 
\hline
95409-01-18-04 & 55327.04 & $4.86^{+0.03}_{-0.02}$ & B & This work & $3.5^{+0.1}_{-0.0}$ & $833^{+20}_{-10}$ & \\ [3pt] 
 & & $4.88^{+0.11}_{-0.07}$ & B & \citet{Gao14} & & & \\ [3pt] 
 & & $4.80^{+0.10}_{-0.10}$ & B & \citet{Motta11} & & & \\ [3pt] 
 & & $4.78^{+0.08}_{-0.12}$ & B & \citet{Nandi12} & & & \\ [3pt] 
\hline
95409-01-18-05 & 55327.26 & $4.95^{+0.02}_{-0.02}$ & B & This work & $3.7^{+0.3}_{-0.2}$ & $774^{+72}_{-69}$ & \\ [3pt] 
 & & $4.91^{+0.03}_{-0.03}$ & B & \citet{Gao14} & & & \\ [3pt] 
 & & $4.90^{+0.04}_{-0.04}$ & B & \citet{Motta11} & & & \\ [3pt] 
 & & $4.95^{+0.04}_{-0.04}$ & B & \citet{Nandi12} & & & \\ [3pt] 
\hline
95409-01-19-00 & 55330.29 & $4.70^{+0.02}_{-0.02}$ & B & This work & $3.5^{+0.1}_{-0.1}$ & $854^{+41}_{-32}$ & \\ [3pt] 
 & & $4.70^{+0.04}_{-0.03}$ & B & \citet{Gao14} & & & \\ [3pt] 
 & & $4.70^{+0.05}_{-0.05}$ & B & \citet{Motta11} & & & \\ [3pt] 
 & & $4.69^{+0.03}_{-0.04}$ & B & \citet{Nandi12} & & & \\ [3pt] 
\hline
96409-01-04-01 & 55584.37 & $1.72^{+0.08}_{-0.05}$ & B & \citet{Nandi12} & $3.0^{+0.3}_{-0.1}$ & $1064^{+145}_{-67}$ & \\ [3pt] 
\hline
96409-01-04-04 & 55585.95 & $2.03^{+0.03}_{-0.03}$ & B & This work & $3.0^{+0.2}_{-0.1}$ & $1090^{+114}_{-55}$ & \\ [3pt] 
 & & $2.06^{+0.06}_{-0.06}$ & B & \citet{Gao14} & & & \\ [3pt] 
 & & $2.00^{+0.10}_{-0.10}$ & B & \citet{Motta11} & & & \\ [3pt] 
 & & $2.05^{+0.16}_{-0.20}$ & B & \citet{Nandi12} & & & \\ [3pt] 
\hline
96409-01-04-05 & 55586.30 & $0.86^{+0.12}_{-0.09}$ & B & \citet{Gao14} & $3.0^{+0.1}_{-0.1}$ & $1090^{+77}_{-41}$ & \\ [3pt] 
 & & $0.90^{+0.10}_{-0.10}$ & B & \citet{Motta11} & & & \\ [3pt] 
\hline
96409-01-04-02 & 55586.49 & $2.19^{+0.08}_{-0.08}$ & B & This work & $3.0^{+0.1}_{-0.1}$ & $1090^{+77}_{-41}$ & \\ [3pt] 
 & & $2.14^{+0.16}_{-0.14}$ & B & \citet{Nandi12} & & & \\ [3pt] 
\hline
96409-01-05-01 & 55591.62 & $1.74^{+0.06}_{-0.05}$ & B & This work & $2.9^{+0.1}_{-0.1}$ & $1117^{+79}_{-42}$ & \\ [3pt] 
 & & $1.79^{+0.10}_{-0.09}$ & B & \citet{Gao14} & & & \\ [3pt] 
 & & $1.70^{+0.10}_{-0.10}$ & B & \citet{Motta11} & & & \\ [3pt] 
 & & $1.71^{+0.12}_{-0.13}$ & B & \citet{Nandi12} & & & \\ [3pt] 
\hline
96409-01-05-05 & 55592.74 & $2.17^{+0.11}_{-0.22}$ & B & \citet{Nandi12} & $3.1^{+0.2}_{-0.1}$ & $1038^{+85}_{-39}$ & \\ [3pt] 
\hline
96409-01-05-02 & 55592.90 & $1.73^{+0.07}_{-0.06}$ & B & \citet{Gao14} & $3.1^{+0.2}_{-0.1}$ & $1038^{+85}_{-39}$ & \\ [3pt] 
 & & $1.80^{+0.10}_{-0.10}$ & B & \citet{Motta11} & & & \\ [3pt] 
 & & $1.74^{+0.08}_{-0.06}$ & B & \citet{Nandi12} & & & \\ [3pt] 
\hline
96409-01-05-03 & 55594.90 & $6.42^{+0.27}_{-0.22}$ & C & \citet{Nandi12} & $3.4^{+0.0}_{-0.0}$ & $896^{+0}_{-0}$ & \\ [3pt] 
\hline
96409-01-06-00 & 55597.27 & $4.68^{+0.41}_{-0.40}$ & C & \citet{Nandi12} & $8.0^{+1.0}_{-1.6}$ & $245^{+41}_{-100}$ & \\ [3pt] 
\hline
96409-01-06-01 & 55598.67 & $4.53^{+0.14}_{-0.15}$ & C & This work & $10.4^{+0.0}_{-0.0}$ & $166^{+0}_{-0}$ & \\ [3pt] 
 & & $4.52^{+0.30}_{-0.30}$ & C & \citet{Motta11} & & & \\ [3pt] 
 & & $3.97^{+0.12}_{-0.16}$ & C & \citet{Nandi12} & & & \\ [3pt] 
\hline
96409-01-06-02 & 55601.89 & $1.32^{+0.02}_{-0.00}$ & C & \citet{Nandi12} & $11.4^{+5.8}_{-4.5}$ & $143^{+66}_{-162}$ & \\ [3pt] 
\hline
96409-01-07-00 & 55604.00 & $1.09^{+0.01}_{-0.01}$ & C & \citet{Nandi12} & $27.6^{+8.2}_{-9.8}$ & $38^{+12}_{-36}$ & \\ [3pt] 
\hline
96409-01-07-03 & 55604.90 & $1.15^{+0.11}_{-0.06}$ & C & \citet{Nandi12} & $39.5^{+15.2}_{-13.2}$ & $22^{+9}_{-19}$ & \hspace{1cm}
\end{supertabular}%

\twocolumn

\begin{table*}
\caption{Newly detected QPOs in \textit{RXTE}/PCA data from \gx.}
\label{tab:newqpos}
\begin{center}
\begin{tabular}{c c c c c c c}
\hline
\hline
ObsID & Time & Frequency & Width & \%rms & Type\\ 
& (MJD) & (Hz) & (Hz) & & & \\ 
\hline
90110-02-01-01 & 53232.01 & $5.05^{+0.06}_{-0.04}$ & $0.28^{+0.19}_{-0.19}$ & $7.11^{+1.79}_{-1.62}$ & C \\ [3pt] 
70110-01-10-00 & 52402.49 & $2.06^{+0.01}_{-0.01}$ & $0.25^{+0.04}_{-0.03}$ & $11.93^{+0.57}_{-0.56}$ & C \\ [3pt] 
70109-01-20-00 & 52504.71 & $4.76^{+0.08}_{-0.09}$ & $1.86^{+0.25}_{-0.22}$ & $3.75^{+0.16}_{-0.16}$ & B \\ [3pt] 
92085-01-03-00 & 54161.67 & $7.72^{+0.18}_{-0.18}$ & $2.69^{+0.55}_{-0.46}$ & $3.21^{+0.23}_{-0.22}$ & B \\ [3pt] 
92085-01-03-03 & 54164.56 & $7.39^{+0.22}_{-0.21}$ & $3.43^{+1.11}_{-0.83}$ & $3.54^{+0.31}_{-0.32}$ & B \\ [3pt] 
95409-01-04-00 & 55225.71 & $2.48^{+0.02}_{-0.02}$ & $0.39^{+0.05}_{-0.05}$ & $15.83^{+0.63}_{-0.60}$ & C \\ [3pt]
95409-01-17-02 & 55318.44 & $6.66^{+0.13}_{-0.12}$ & $0.97^{+0.58}_{-0.42}$ & $5.00^{+4.73}_{-1.00}$ & C \\ [3pt] 
\hline
\end{tabular}
\end{center}
\end{table*}

\end{document}